\documentclass[twocolumn]{aastex61}  

\usepackage{mathtools,hyperref}
\usepackage{amsmath}
\usepackage{graphicx}
\usepackage{natbib}
\usepackage[all]{hypcap}								
\usepackage{xcolor}
\usepackage{hyperref}
\usepackage{braket}
\usepackage{fancyhdr}
\usepackage{bm}
\hypersetup{linkcolor=cyan,colorlinks=true,filecolor=cyan,citecolor=cyan}

\def\tocm{$21\,\textrm{cm}$\ }

\def\kpara{k_{\parallel}}
\def\kperp{k_{\perp}}

\begin{document}

\title{Mitigating Internal Instrument Coupling for 21 cm Cosmology II:\\ A Method Demonstration with the Hydrogen Epoch of Reionization Array}
\shorttitle{Mitigating Internal Instrument Coupling for 21 cm Cosmology II}
\shortauthors{Kern et al.}

\correspondingauthor{Nicholas Kern}
\email{nkern@berkeley.edu}
\author{Nicholas S. Kern}
\affiliation{Department of Astronomy, University of California, Berkeley, CA}

\author{Aaron R. Parsons}
\affiliation{Department of Astronomy, University of California, Berkeley, CA}

\author{Joshua S. Dillon}
\affiliation{Department of Astronomy, University of California, Berkeley, CA}
\affiliation{NSF Astronomy and Astrophysics Postdoctoral Fellow}

\author{Adam E. Lanman}
\affiliation{Department of Physics, Brown University, Providence, RI}

\author{Adrian  Liu}
\affiliation{Department of Physics and McGill Space Institute, McGill University, Montreal, Canada}

\author{Philip Bull}
\affiliation{School of Physics \& Astronomy, Queen Mary University of London, London, UK}

\author{Aaron  Ewall-Wice}
\affiliation{Jet Propulsion Laboratory, California Institute of Technology, Pasadena, CA}

\author{Zara  Abdurashidova}
\affiliation{Department of Astronomy, University of California, Berkeley, CA}

\author{James E. Aguirre}
\affiliation{Department of Physics and Astronomy, University of Pennsylvania, Philadelphia, PA}

\author{Paul  Alexander}
\affiliation{Cavendish Astrophysics, University of Cambridge, Cambridge, UK}

\author{Zaki S. Ali}
\affiliation{Department of Astronomy, University of California, Berkeley, CA}

\author{Yanga  Balfour}
\affiliation{SKA-SA, Cape Town, South Africa}

\author{Adam P. Beardsley}
\affiliation{School of Earth and Space Exploration, Arizona State University, Tempe, AZ}
\affiliation{NSF Astronomy and Astrophysics Postdoctoral Fellow}

\author{Gianni  Bernardi}
\affiliation{Department of Physics and Electronics, Rhodes University, PO Box 94, Grahamstown, 6140, South Africa}
\affiliation{INAF-Istituto di Radioastronomia, via Gobetti 101, 40129 Bologna, Italy}
\affiliation{SKA-SA, Cape Town, South Africa}

\author{Judd D. Bowman}
\affiliation{School of Earth and Space Exploration, Arizona State University, Tempe, AZ}

\author{Richard F. Bradley}
\affiliation{National Radio Astronomy Observatory, Charlottesville, VA}

\author{Jacob  Burba}
\affiliation{Department of Physics, Brown University, Providence, RI}

\author{Chris L. Carilli}
\affiliation{National Radio Astronomy Observatory, Socorro, NM}

\author{Carina  Cheng}
\affiliation{Department of Astronomy, University of California, Berkeley, CA}

\author{David R. DeBoer}
\affiliation{Department of Astronomy, University of California, Berkeley, CA}

\author{Matt  Dexter}
\affiliation{Department of Astronomy, University of California, Berkeley, CA}

\author{Eloy  de~Lera~Acedo}
\affiliation{Cavendish Astrophysics, University of Cambridge, Cambridge, UK}

\author{Nicolas  Fagnoni}
\affiliation{Cavendish Astrophysics, University of Cambridge, Cambridge, UK}

\author{Randall  Fritz}
\affiliation{SKA-SA, Cape Town, South Africa}

\author{Steve R. Furlanetto}
\affiliation{Department of Physics and Astronomy, University of California, Los Angeles, CA}

\author{Brian  Glendenning}
\affiliation{National Radio Astronomy Observatory, Socorro, NM}

\author{Deepthi  Gorthi}
\affiliation{Department of Astronomy, University of California, Berkeley, CA}

\author{Bradley  Greig}
\affiliation{School of Physics, University of Melbourne, Parkville, VIC 3010, Australia}
\affiliation{ ARC Centre of Excellence for All-Sky Astrophysics in 3 Dimensions
(ASTRO 3D), University of Melbourne, VIC 3010, Australia}

\author{Jasper  Grobbelaar}
\affiliation{SKA-SA, Cape Town, South Africa}

\author{Ziyaad  Halday}
\affiliation{SKA-SA, Cape Town, South Africa}

\author{Bryna J. Hazelton}
\affiliation{Department of Physics, University of Washington, Seattle, WA}
\affiliation{eScience Institute, University of Washington, Seattle, WA}

\author{Jacqueline N. Hewitt}
\affiliation{Department of Physics, Massachusetts Institute of Technology, Cambridge, MA}

\author{Jack  Hickish}
\affiliation{Department of Astronomy, University of California, Berkeley, CA}

\author{Daniel C. Jacobs}
\affiliation{School of Earth and Space Exploration, Arizona State University, Tempe, AZ}

\author{Austin  Julius}
\affiliation{SKA-SA, Cape Town, South Africa}

\author{Joshua  Kerrigan}
\affiliation{Department of Physics, Brown University, Providence, RI}

\author{Piyanat  Kittiwisit}
\affiliation{School of Earth and Space Exploration, Arizona State University, Tempe, AZ}

\author{Saul A. Kohn}
\affiliation{Department of Physics and Astronomy, University of Pennsylvania, Philadelphia, PA}

\author{Matthew  Kolopanis}
\affiliation{School of Earth and Space Exploration, Arizona State University, Tempe, AZ}

\author{Paul  La~Plante}
\affiliation{Department of Physics and Astronomy, University of Pennsylvania, Philadelphia, PA}

\author{Telalo  Lekalake}
\affiliation{SKA-SA, Cape Town, South Africa}

\author{David  MacMahon}
\affiliation{Department of Astronomy, University of California, Berkeley, CA}

\author{Lourence  Malan}
\affiliation{SKA-SA, Cape Town, South Africa}

\author{Cresshim  Malgas}
\affiliation{SKA-SA, Cape Town, South Africa}

\author{Matthys  Maree}
\affiliation{SKA-SA, Cape Town, South Africa}

\author{Zachary E. Martinot}
\affiliation{Department of Physics and Astronomy, University of Pennsylvania, Philadelphia, PA}

\author{Eunice  Matsetela}
\affiliation{SKA-SA, Cape Town, South Africa}

\author{Andrei  Mesinger}
\affiliation{Scuola Normale Superiore, 56126 Pisa, PI, Italy}

\author{Mathakane  Molewa}
\affiliation{SKA-SA, Cape Town, South Africa}

\author{Miguel F. Morales}
\affiliation{Department of Physics, University of Washington, Seattle, WA}

\author{Tshegofalang  Mosiane}
\affiliation{SKA-SA, Cape Town, South Africa}

\author{Steven G. Murray}
\affiliation{School of Earth and Space Exploration, Arizona State University, Tempe, AZ}

\author{Abraham R. Neben}
\affiliation{Department of Physics, Massachusetts Institute of Technology, Cambridge, MA}

\author{Aaron R. Parsons}
\affiliation{Department of Astronomy, University of California, Berkeley, CA}

\author{Nipanjana  Patra}
\affiliation{Department of Astronomy, University of California, Berkeley, CA}

\author{Samantha  Pieterse}
\affiliation{SKA-SA, Cape Town, South Africa}

\author{Jonathan C. Pober}
\affiliation{Department of Physics, Brown University, Providence, RI}

\author{Nima  Razavi-Ghods}
\affiliation{Cavendish Astrophysics, University of Cambridge, Cambridge, UK}

\author{Jon  Ringuette}
\affiliation{Department of Physics, University of Washington, Seattle, WA}

\author{James  Robnett}
\affiliation{National Radio Astronomy Observatory, Socorro, NM}

\author{Kathryn  Rosie}
\affiliation{SKA-SA, Cape Town, South Africa}

\author{Peter  Sims}
\affiliation{Department of Physics, Brown University, Providence, RI}

\author{Craig  Smith}
\affiliation{SKA-SA, Cape Town, South Africa}

\author{Angelo  Syce}
\affiliation{SKA-SA, Cape Town, South Africa}

\author{Nithyanandan  Thyagarajan}
\affiliation{School of Earth and Space Exploration, Arizona State University, Tempe, AZ}
\affiliation{National Radio Astronomy Observatory, Socorro, NM}
\affiliation{Jansky Fellow}

\author{Peter K.~G. Williams}
\affiliation{Harvard-Smithsonian Center for Astrophysics, Cambridge, MA}

\author{Haoxuan  Zheng}
\affiliation{Department of Physics, Massachusetts Institute of Technology, Cambridge, MA}

\begin{abstract}
\newpage
We present a study of internal reflection and cross coupling systematics in Phase 1 of the Hydrogen Epoch of Reionization Array (HERA).
In a companion paper, we outlined the mathematical formalism for such systematics and presented algorithms for modeling and removing them from the data.
In this work, we apply these techniques to data from HERA's first observing season  as a method demonstration.
The data show evidence for systematics that, without removal, would hinder a detection of the 21 cm power spectrum for the targeted EoR line-of-sight modes in the range $0.2 < \kpara < 0.5\ h^{-1}$ Mpc.
After systematic removal, we find we can recover these modes in the power spectrum down to the integrated noise-floor of a nightly observation, achieving a dynamic range in the EoR window of $10^{-6}$ in power (mK$^2$ units) with respect to the bright galactic foreground signal.
In the absence of other systematics and assuming the systematic suppression demonstrated here continues to lower noise levels, our results suggest that fully-integrated HERA Phase I may have the capacity to set competitive upper limits on the 21 cm power spectrum.
For future observing seasons, HERA will have upgraded analog and digital hardware to better control these systematics in the field.
\end{abstract}

\defcitealias{Kern2019a}{K19a}


\section{Introduction}
\label{sec:intro}
The Epoch of Reionization (EoR) marks a fundamental phase transition in cosmic history, where neutral hydrogen filling the intergalactic medium (IGM) was ionized by a radiation field thought to originate from the formation of the first generation of stars and galaxies in the universe \citep[for reviews, see][]{Furlanetto2006c, Loeb2013, Mesinger2016b}.
One of the only direct probes of the IGM throughout the entirety of the EoR is neutral hydrogen's \tocm line.
A hyperfine transition of neutral hydrogen, \tocm emission is a three dimensional, tomographic probe of the IGM's density, ionization and temperature structure.
Low-frequency radio surveys promise to revolutionize our understanding of the IGM by using the \tocm line to map out its morphology during EoR, and place constraints on the sources responsible for its heating and eventual reionization.

Over the past decade, experiments like the Donald C. Backer Precision Array for Probing the Epoch of Reionization \citep[PAPER;][]{Parsons2014, Jacobs2015, Ali2015}, the Murchison Widefield Array \citep[MWA;][]{Dillon2014, Ewall-Wice2016b, Beardsley2016}, the Low Frequency Array \citep[LOFAR;][]{Patil2017}, and the Giant Metre Wave Radio Telescope \citep[GMRT;][]{Paciga2013} have placed increasingly competitive limits on the \tocm power spectrum.
These experiments face the challenge of separating a weak cosmological signal from foreground emission that is generally $10^5$ times brighter in order to characterize the EoR.
Instrumental systematics further complicate this effort, which can cause foreground signal to contaminate Fourier modes in the data that would otherwise only be noise limited.
As such, many of the current upper limits on the \tocm power spectrum have been limited by instrumental systematics.
Current and future experiments like the Hydrogen Epoch of Reionization Array \citep[HERA;][]{DeBoer2017} and the Square Kilometer Array \citep[SKA;][]{Koopmans2015} are nominally forecasted to provide high significance characterizations of the \tocm signal and place constraints on IGM properties and the sources driving reionization \citep{Pober2014, Greig2015a, Greig2015b, Liu2016b, Ewall-Wice2016b, Greig2017b, Kern2017}.
However, these forecasts neglect the impact of systematic contamination, which can significantly hamper an experiment's overall sensitivity and parameter constraining ability.
Precise modeling and separation of instrumental systematics will therefore likely be necessary for second-generation \tocm experiments to make robust detections of the cosmological \tocm signal.

Systematic contamination comes in a variety of forms, including calibration errors, ionospheric faraday rotation, primary beam ellipticity, analogue signal chain imperfections (such as impedance mismatches), and others.
In a companion paper, \citet{Kern2019a}, we presented techniques for modeling and removing systematics specifically due to internal instrument coupling, such as signal chain reflections and antenna cross coupling (i.e. crosstalk).
In that paper, we describe the phenomenology of internal instrument systematics in the interferometric visibilities, propose algorithms for removing them from the data, and demonstrate their performance against numerical simulations.
In this work, we investigate data from HERA Phase I for internal instrument systematics and apply our systematic modeling algorithms as a proof-of-concept.

The structure of this paper is as follows.
In \S2 we describe the data and observations used for this analysis.
In \S3 we examine the data for signal chain reflections,
and demonstrate reflection calibration on HERA auto-correlation visibilities.
In \S4 we present a study of cross coupling systematics in the HERA system, and demonstrate cross coupling removal performance on a few select baselines.
In \S5 we perform joint reflection and cross coupling systematic removal for baselines across the entire HERA array and compute power spectra, and lastly in \S6 we summarize our results.

\section{Observations}
\label{sec:hera_data}
\begin{figure}
\centering
\label{fig:sigchain}
\includegraphics[scale=0.3]{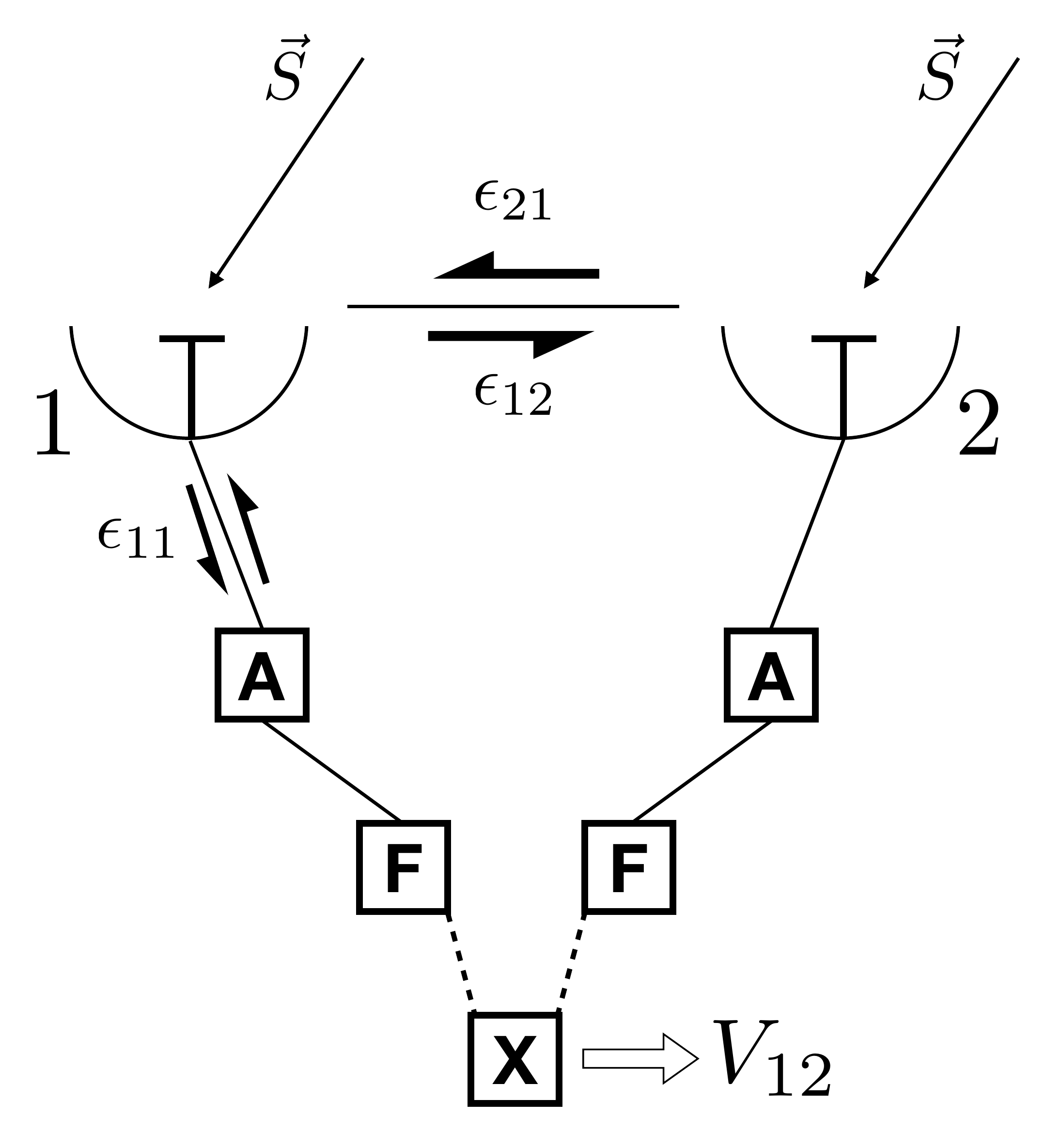}
\caption{A schematic of HERA signal chains for two antennas, 1 \& 2.
Sky signal ($\vec{\mathcal{S}}$) enters each antenna's dish and feed where it is converted into a voltage, travels down a 150-m coaxial cable to a processing node holding a post-amplification module ($\mathbf{A}$), before being directed through a 20-m cable to an engine that digitizes and Fourier transforms the signal ($\mathbf{F}$) and then sent to the correlator ($\mathbf{X}$) to produce the visibility $V_{12}$.
A possible cable reflection in antenna 1's signal chain is marked as $\epsilon_{11}$, traversing up and down the cable connecting the feed to the node.
A possible source of feed-to-feed coupling is marked as $\epsilon_{12}$ and $\epsilon_{21}$, where a signal is reflected off of antenna 1's feed and into antenna 2's feed or vice versa.
The dashed line from $\mathbf{F}$ to $\mathbf{X}$ denotes a signal pathway after digitization, where reflection are not a concern.}
\end{figure}

Data were taken during HERA Phase I, which observed from 2017 to 2018 while undergoing active construction \citep{DeBoer2017}.
The Phase I instrument was a hybrid HERA-PAPER system, taking the signal chains and correlator from the PAPER experiment \citep{Parsons2010, Ali2015} and attaching them to new HERA antennas.
The HERA antenna is a parabolic dish spanning 14 meters in diameter, with a focal height designed to minimize reflections within the dish \citep{Neben2016, Thyagarajan2016, Ewall-Wice2016c, Patra2018}.
The feed uses the PAPER sleeved-dipole as the active element, which in the Phase I instrument has been optimized for the HERA antenna \citep{DeBoer2015, Fagnoni2016}.
An active balun or front-end module (FEM) is connected to the feed and houses a low-noise amplifier.
After initial amplification, the signals are sent through a 150-meter coaxial cable (first cable in \autoref{fig:sigchain}) to a node unit in the field holding a post-amplifier module (PAM; A box in \autoref{fig:sigchain}), and are then sent through another coaxial cable of about 20 meters in length (second cable in \autoref{fig:sigchain}) to a container holding ROACH2 boards\footnote{\url{https://casper.ssl.berkeley.edu/wiki/ROACH2}} \citep{Parsons2008, Hickish2016} that digitize the signals and then Fourier transforms them into the frequency domain (F box in \autoref{fig:sigchain}).
Finally, a Graphics Processing Unit (GPU) correlator cross-multiplies the signals between all antenna pairs to form interferometric visibilities that are integrated for 10.7 seconds before being written to disk (X box in \autoref{fig:sigchain}).

\begin{figure}
\centering
\label{fig:array_layout}
\includegraphics[scale=0.45]{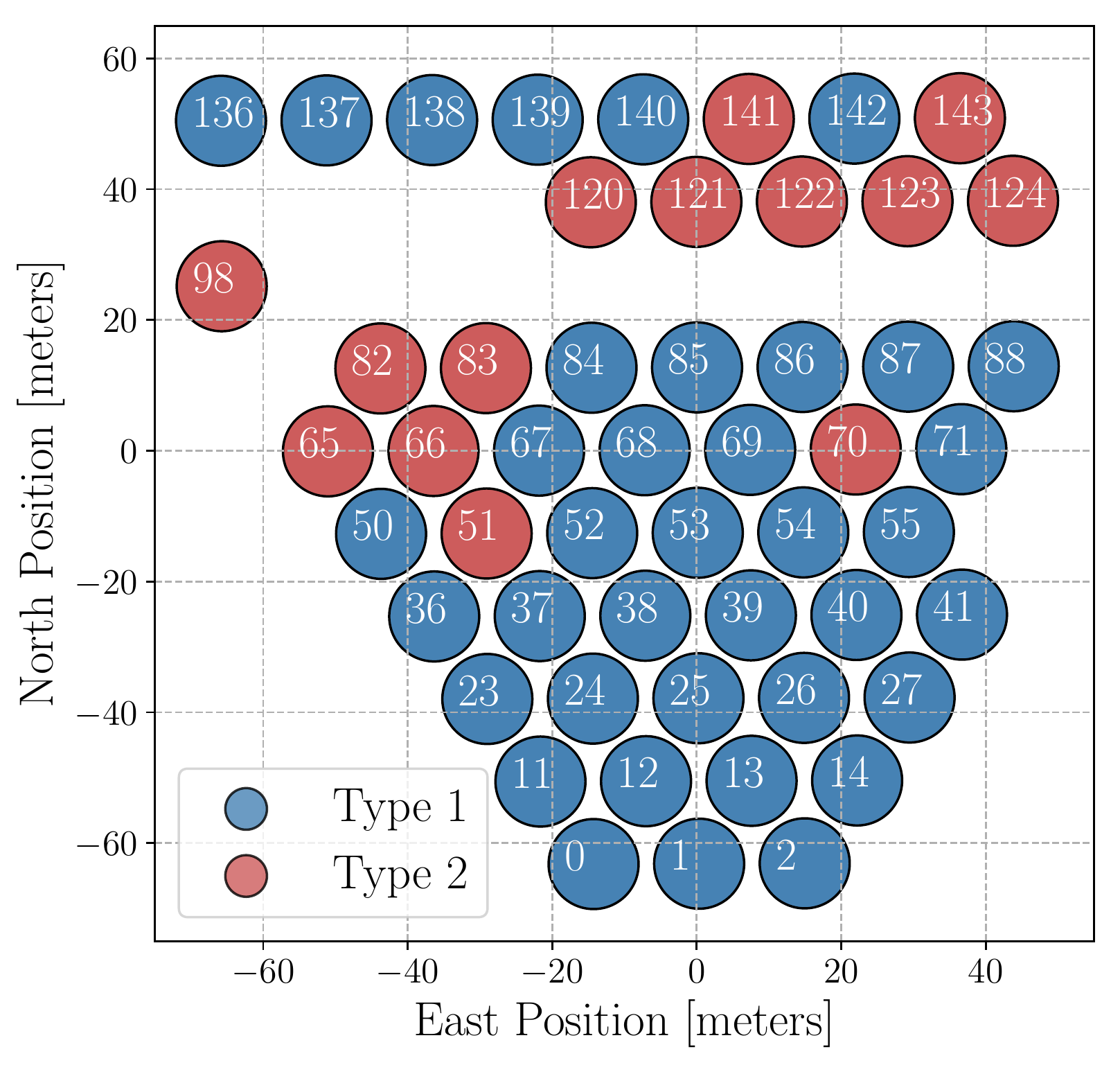}
\caption{The HERA array configuration at the time of observations on Julian Date 2458101 with roughly 46 operational antennas, showing which fall into Type 1 (blue) and Type 2 (red) signal chains categories.}
\end{figure}

\begin{deluxetable}{lc} 
\tabletypesize{\footnotesize} 
\tablewidth{0pt} 
\tablecaption{
HERA Observation Parameters
\label{tab:hera_obs}
}
\tablehead{Parameter & Value}
\startdata 
Observation Date & December 13, 2017 \\[.1cm]
Array Coordinates  & -30.7$^\circ$ S, 21.4$^\circ$E  \\[.1cm]
JD Range & 2458101.27 - 2458101.61 \\[.1cm]
LST Range &  1.5 - 9.6 hours \\[.1cm]
Integration Time & 10.7 seconds \\[.1cm]
Frequency Range & 100 - 200 MHz \\[.1cm]
Channel Width & 97.65 kHz \\[.1cm]
Dish Diameter & 14 meter \\[.1cm]
Feed Type & PAPER dipole \\[.1cm]
Instrumental Polarization & North-South (``YY'') \\[.1cm]
Cable Type & 150-m \& 20-m coaxial \\[.1cm]
\enddata 
\tablecomments{For the 2017--2018 observation, the HERA correlator used the convention that the X dipole points East-West while the Y dipole points North-South, which is not the standard \citet{Hamaker1996b} definition.}
\end{deluxetable}

The observations presented in this work come from a single night spanning 8 hours of local sidereal time (LST) on Julian Date 2458101.
At that time, the array consisted of 46 operational antennas, each with dual-polarization dipole feeds (\autoref{fig:array_layout}).
Additionally, the signal chains of the array were split into two categories: Type 1 which used newly manufactured FEMs, PAMs, and coaxial cables specifically for HERA Phase I, and Type 2 which re-purposed the PAPER FEMs, PAMs and coaxial cables (colored blue and red in \autoref{fig:array_layout}, respectively).
In this analysis, we only use North-South (``YY'') linear dipole polarization data, although all four auto and cross-feed polarization data products are recorded by the correlator.
Additional observational parameters are tabulated in \autoref{tab:hera_obs}.

\begin{figure}
\centering
\label{fig:flag_mask}
\includegraphics[scale=0.55]{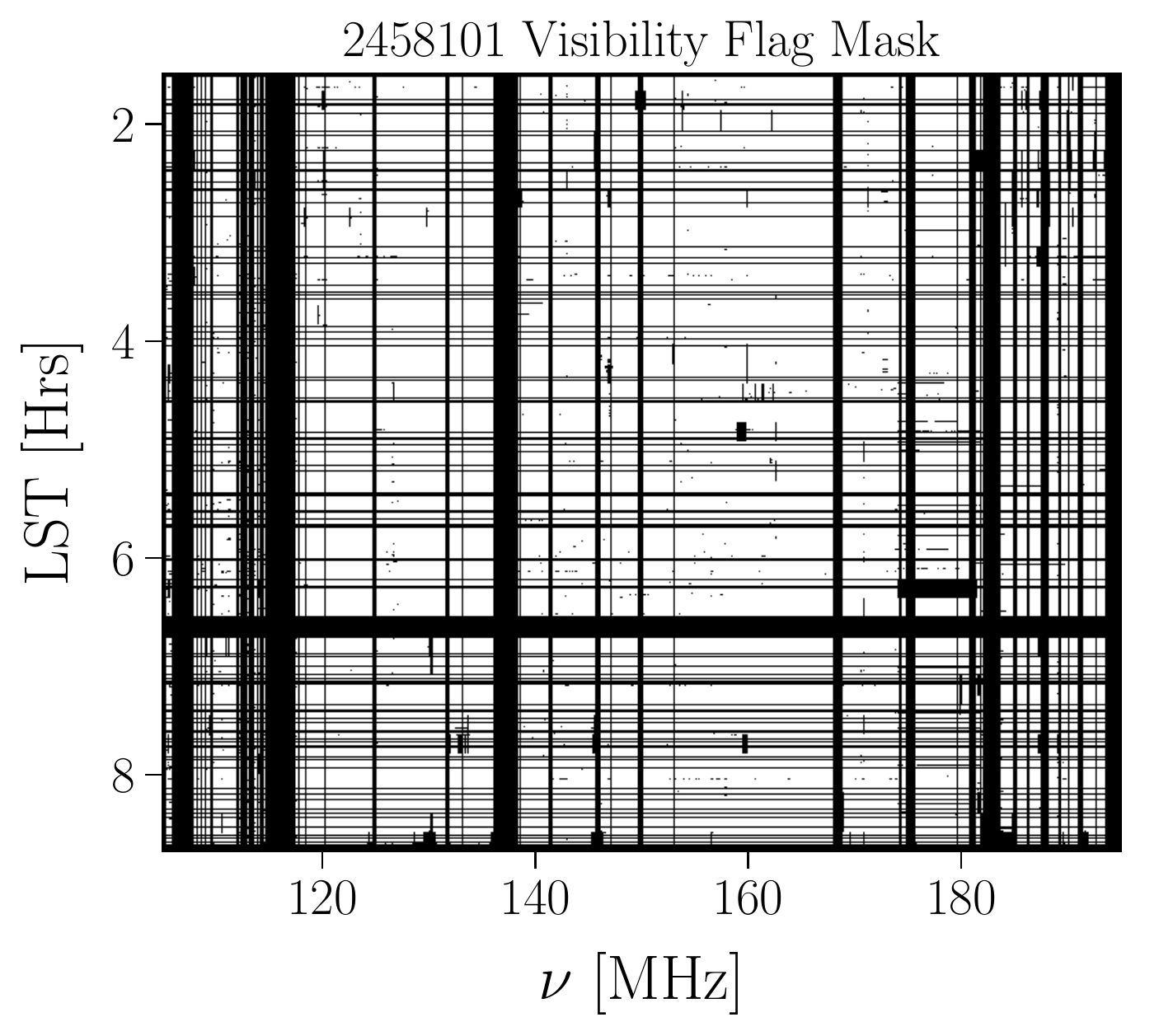}
\caption{Aggressive RFI visibility mask as a function of time and frequency after three rounds of flagging.
Flags are broadcasted across time and/or frequency if a 15\% flagged threshold is met per time bin and frequency channel.
For this particular night $\sim30$\% of the total data volume is flagged, which is sub-optimal in that it is likely a significant over-flagging,
but with the benefit of being more aggressive in flagging low-level and repeating RFI.}
\end{figure}

The data have been pre-processed with part of the HERA reduction and calibration pipeline.
Specifically, the data are first flagged for radio frequency interference (RFI) using a median filter and watershed algorithm operating on the cross correlation visibilities \citep{Kerrigan2019, Beardsley2019}.
In this work, we also enact two additional steps for RFI flagging.
The first takes stacked auto-correlation visibilities and differences them across time and frequency, normalizes them by their median absolute deviation and flags the residual at the 4 sigma level.
Our second step runs a delay-based, iterative deconvolution on a subset of the auto-correlation visibilities, which attempts to deconvolve the discontinuous windowing function created by flagged data.
This is similar in concept to the image-based CLEAN deconvolution \citep{Hogbom1974}, except applied to the frequency and delay domains rather than the $uv$ and $lm$ domains, and with the missing data coming from RFI rather than incomplete $uv$ sampling.
We then normalize the filtered residual in frequency space by its median absolute deviation, and again enact RFI cuts at the 4 sigma level.
Flags from each of the three independent steps are combined with a logical OR and then broadcasted across time and/or frequency if a 15\% flagged threshold is met for any individual time bin or frequency channel.
An example of the fairly aggressive resultant visibility flagging mask is shown in \autoref{fig:flag_mask}.
In total roughly 30\% of the data volume is flagged, although this likely contains a decent amount of over-flagging.

Next we calibrate the data using a highly simplified antenna-based calibration.
The full HERA calibration pipeline computes complex antenna gains for each time integration over the entire night from a combination of redundant calibration \citep{Dillon2019} and a constrained absolute calibration with the resultant gains smoothed across time and frequency \citep{Kern2019c}.
In this work, we take the gains derived from these steps and 1) average them across the entire night into a single spectrum, 2) average their amplitude across frequency to a single number, and 3) fit for a phase-slope across frequency (i.e. a single antenna delay).
We are left with a single amplitude and delay for each antenna, which we apply to all times of the night.
This has the effect of properly setting the flux scale of the data and also calibrates out the antenna cable delay, but ensures the gain itself we apply to the data has little to no spectral structure.

Because of our highly simplified calibration, the instrumental bandpass is not corrected for and still exists in the data.
Calibration, being multiplicative in frequency space, can be thought of as a convolution in delay space.
The true response of the visibilities in delay space is therefore initially convolved by the bandpass kernel upon measurement by the telescope.
Assuming the bandpass is composed primarily of large-scale modes, its impact will be a slight smoothing-out of the true sky delay response and features created by systematics.
Bandpass calibration performed beforehand may therefore sharpen systematics in delay space and actually make it easier to model and remove them.
Antenna-based calibration for HERA in the context of redundant calibration and absolute calibration is explored in \citet{Dillon2019} and \citet{Kern2019c}.


\begin{figure*}
\centering
\label{fig:autoamp_avg}
\includegraphics[scale=0.55]{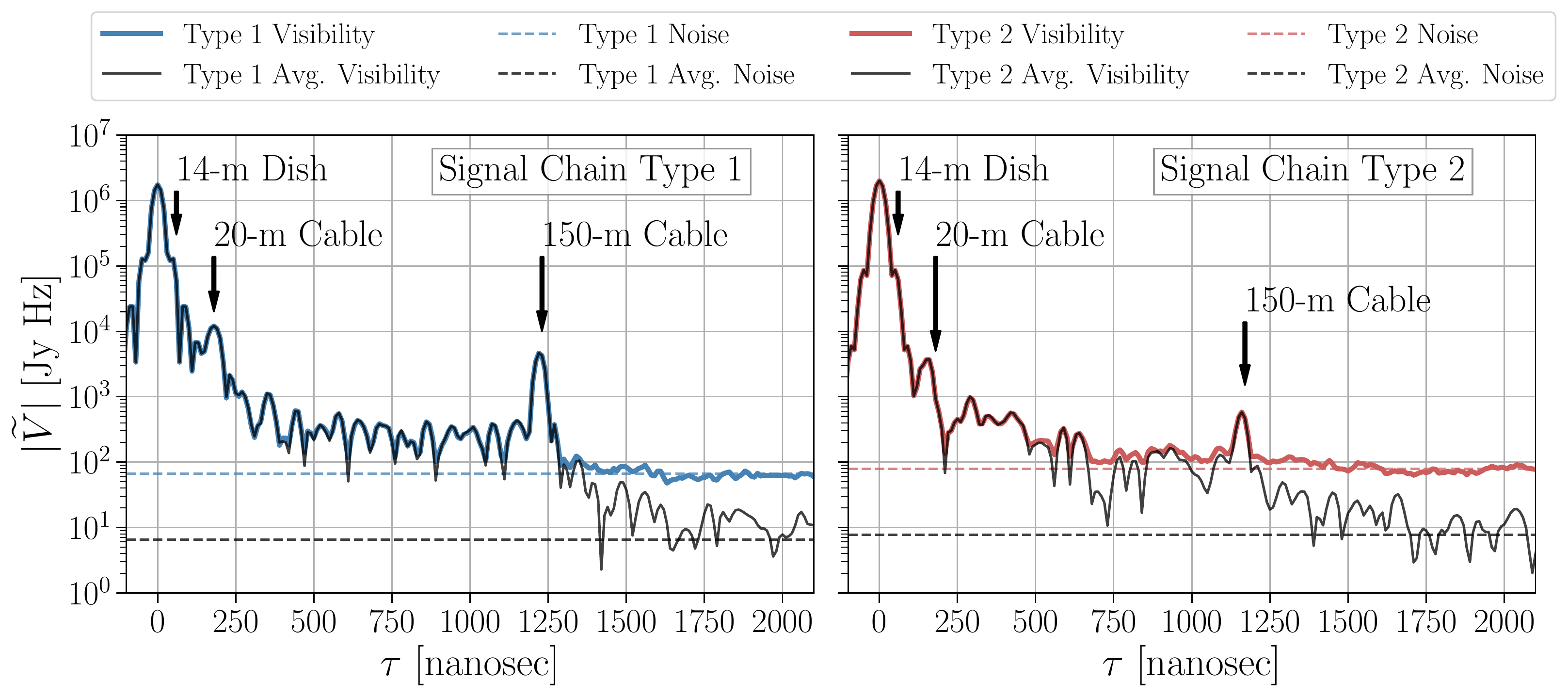}
\caption{Auto-correlation visibilities for signal chain Type 1 (left) and Type 2 (right) with absolute time averaging (blue \& red) and with complex time averaging (black), and their associated noise floors (dashed).
Antennas 84 and 121 were used for the two auto-correlation visibilities.
Delays for relevant length scales in the analogue system are marked with arrows.
Resonances in the dish and reflections in the cables tend to be worse for signal chain Type 1.
Additionally, we see evidence for a systematic tail in both signal chain types spanning a wide range of delays that does not integrate down like noise.}
\end{figure*}

\section{Signal Chain Reflections}
\label{sec:ref_cal}
In this section we inspect the data for evidence of signal chain reflections.
To do this, we take the auto-correlation visibility from each antenna and look for peaks in delay space (see \citealt{Kern2019a} for a summary of the algorithm).
The calibrated data are filled with flags due to RFI (\autoref{fig:flag_mask}) and are thus nulled to zero at the flagged channels.
This is not ideal for inspecting the data in delay space, as the Fourier transform of such a discontinuous windowing function
creates strong sidelobes.
To mitigate this we employ the same delay-based, iterative deconvolution algorithm from before to subtract these sidelobes, effectively interpolating across the nulled gaps in the data due to RFI \citep{Parsons2009}.
We allow the deconvolution to place model components out to delays of $|\tau| < 1600$ ns, and iterate until the process reaches 5$\times$ the noise floor of the data.
We then make a copy of the data, and with the first copy we average the absolute value of the deconvolved visibilities in delay space across a few hours of LST.
With the second copy we average the full complex-valued, deconvolved visibilities across the same time range, which will have a lower noise floor due to the complex average.

\autoref{fig:autoamp_avg} shows these data products for the Type 1 (left) and Type 2 (right) signal chain, with the absolute time-averaged data shown in blue or red, and the complex time-averaged data shown in solid black.
Additionally, the thermal noise floors of each data product is plotted as dashed lines, which is estimated from the data via adjacent time and frequency differencing, and then divided by $1/\sqrt{N_{\rm avg}}$ where $N_{\rm avg}$ is the number of complex averages performed on the data.
We find that Type 2 signal chains achieve a better overall impedance match with the analogue system, leading to slightly less structure in the auto-correlations across a wide range of delays.
Nonetheless, we do see evidence for reflections from both the 20-meter and 150-meter cables, with reflection amplitudes in the range of roughly $3\times10^{-3}$ and 1$\times10^{-3}$, respectively.
Of major concern is the tail of the auto-correlation response, which starts at low delays and slopes down to the noise floor out to the 150-meter cable delay.
This tail is over an order of magnitude larger than that predicted by simulations of the HERA dish and feed \citep{Ewall-Wice2016c}.

\begin{figure*}
\centering
\label{fig:refcal_amps}
\includegraphics[scale=0.52]{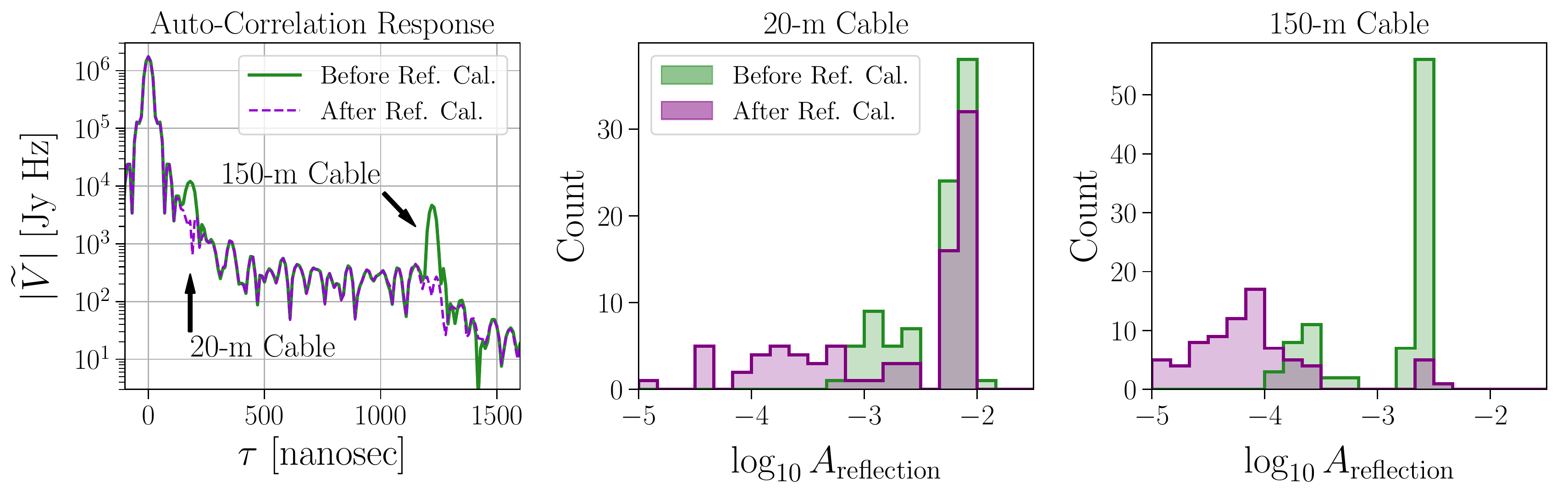}
\caption{Reflection calibration performed over the full band (120 -- 180 MHz) and applied to the auto-correlation visibilities. {\bfseries Left}: The auto-correlation response before calibration (green) and after calibration (purple) demonstrates suppression of reflection systematics by roughly an order of magnitude in the visibility. {\bfseries Center}: Histogram of derived 20-m reflection amplitudes before and after calibration. In the majority of cases we only see suppression by a factor of a few. {\bfseries Right}: Histogram of derived 150-m reflection amplitudes before and after calibration. In the majority of cases we see suppression by at least an order of magnitude.
Less suppression for the 20-m cable is likely attributable to more significant frequency evolution in the reflection parameters.}
\end{figure*}

\begin{figure*}
\centering
\label{fig:autoamp_avg_subbands}
\includegraphics[scale=0.55]{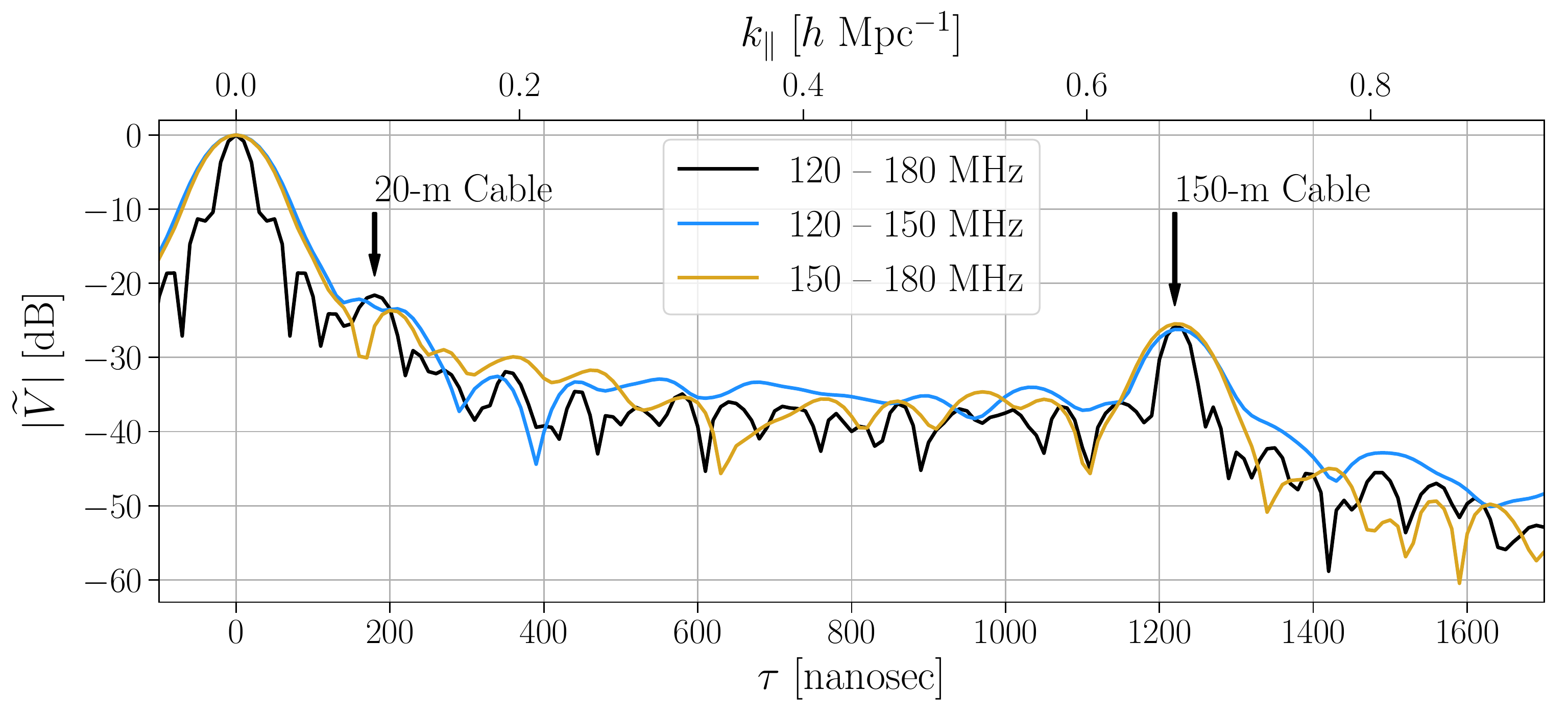}
\caption{Auto-correlation visibility after complex time-averaging, transformed over the full band (120--180 MHz; black), just the low side of the band (120--150 MHz; blue) and just the high side of the band (150--180 MHz; gold) for a Type 1 signal chain.
The 150-m cable reflection parameters are fairly consistent between both sides of the band, while the 20-m cable reflection shows significantly more frequency evolution.
The smaller peaks along the systematic tail also shows significant frequency evolution.
}
\end{figure*}

In this case the noise floor has been integrated down (solid black), we see that delays outside the 150-meter cable delay seem to effectively integrate down with the noise, while delays inside the 150-m cable-delay do not.
This means that the features at low and intermediate delays are coherent on long timescales of at least a few hours.
The abrupt change at $\sim1250$ nanoseconds is also possibly suggestive that tailed response might in part be originating within the 150-m cable.
A possible mechanism for this could be sub-reflections within the cable due to intrinsic cable imperfections or environmental wear and damage along the cable.
Another explanation is the effect of cross coupling (or mutual coupling) between neighboring antennas, which we explore in more detail in cross-correlation visibilities in the following section.
It is not easy to distinguish between these two effects in the auto-correlation visibilities alone.
Direct electromagnetic simulations of mutual coupling in the HERA system provide mixed evidence: predicting it to appear at a similar amplitude and slope in the auto-correlations, but also predicting it to truncate at lower delays of $\sim$600 ns \citep{Fagnoni2019}.

The fact that the auto-correlations show a systematic tail that, for $\tau > 300$ ns  or $k_\parallel > 0.2$ $h$ Mpc$^{-1}$, shows only three to four orders of magnitude of dynamic range is concerning, given that fiducial EoR amplitudes are generally assumed to lie at or below five orders of magnitude in dynamic range in the visibility for similar $k$ \citep{Thyagarajan2016}.
Furthermore, the observed systematic tail extends over a wide range of delays that covers essentially all of the $k_\parallel$ modes of interest ($0.2 < k < 0.6$ $h$ Mpc$^{-1}$).
These systematics need to be well-understood and mitigated if the data are to be used for stringent EoR limits. 

Next we attempt to model some of these features and calibrate them out.
One needs to proceed carefully when doing this because calibrating out structure that is inherent to the true data will actually \emph{create} systematics.
To be conservative, we only target the two features that we know to correlate with the expected delays of the 20-m and 150-m coaxial cables at $\sim200$ and $\sim1250$ ns.
We use the method described in \citet{Kern2019a} to derive reflection parameters across the full bandwidth excluding the band edges (120 -- 180 MHz) and then apply them to the data in frequency space.
\autoref{fig:refcal_amps} shows the result, demonstrating the delay response of an auto-correlation before (green) and after (purple) reflection calibration, and also showing the derived reflection amplitudes of the 20-m and 150-m cable reflections before and after calibration.
We find that in general we can suppress the 150-m cable reflection by a couple orders of magnitude (in the visibility), whereas for the 20-m cable reflection we get on average only a factor of a few suppression.
This may not be enough to remove them below fiducial EoR levels, which is highly concerning for the ultimate performance of the HERA Phase I system.
However, a way to achieve more suppression in the power spectrum is to utilize the highly redundant nature of the array and cross-correlate different baselines of the same orientation when forming power spectra.

Often a limiting factor in reflection modeling is frequency evolution of the reflection parameters \citep{Ewall-Wice2016b}.
In \autoref{fig:autoamp_avg_subbands} we plot the auto-correlation response having taken the Fourier transform of the data over a low-band (120--150 MHz; blue) and a high-band (150--180 MHz; gold), plotted in decibels relative to their peak value.
We observe non-negligible amounts of frequency evolution in the general structure of the systematic tail, with slight evolution for the 150-m cable bump and more significant evolution in the 20-m cable bump.
This is likely at least part of the reason why we achieve less suppression for the 20-m cable reflection,
and suggests that to mitigate reflections to higher dynamic range we will need to perform reflection calibration at the sub-band level.
Because we find the suppression achieved by modeling these reflection across the full band is sufficient for this analysis (\autoref{sec:power_spectra}), we defer sub-band reflection modeling to future studies.

An immediate concern one might have about this technique is the fact that we are applying a calibration with spectral structure at the same or similar delays we hope to use for measuring the EoR power spectrum \citep{Mouri2019}, which may lead to signal loss \citep{Cheng2018}.
In our companion paper we study signal loss in this same scenario with simulated reflection systematics, and we find that although the auto-correlation visibilities may sustain low levels of signal loss, the cross-correlation visibilities show resistance to signal loss across all delays \citep{Kern2019a}.
This is in part due to the subspace that reflection calibration spans relative to the EoR signal: reflection calibration spans an direction-independent, antenna-based space, while EoR is fundamentally a baseline-dependent measurement.
As such, it is hard for reflection calibration to soak up and calibrate out EoR signal from the cross-correlation visibilities.
This is further compounded by the fact that our reflection calibration method only uses the auto-correlation visibilities to derive reflection parameters.
We refer the reader to \citet{Kern2019a} for a more detailed description of the algorithm and our signal loss simulations.


\section{Antenna Cross Couplings}
\label{sec:cross_coupling}
Next we turn our attention to HERA's cross-correlation visibilities in order to probe for antenna cross coupling systematics.
Specifically, we look at the North-South instrumental polarization (also denoted as `YY') for baselines  (11, 12), (11, 13) \& (11, 14), which are three East-West baselines with lengths of 15, 29 and 44 meters, respectively (\autoref{fig:array_layout}).
These baselines display some of the strongest cross coupling systematics seen in the data, but are otherwise fairly nominal baselines.

\begin{figure}
\centering
\label{fig:cross_corr_freq_spectra}
\includegraphics[scale=0.5]{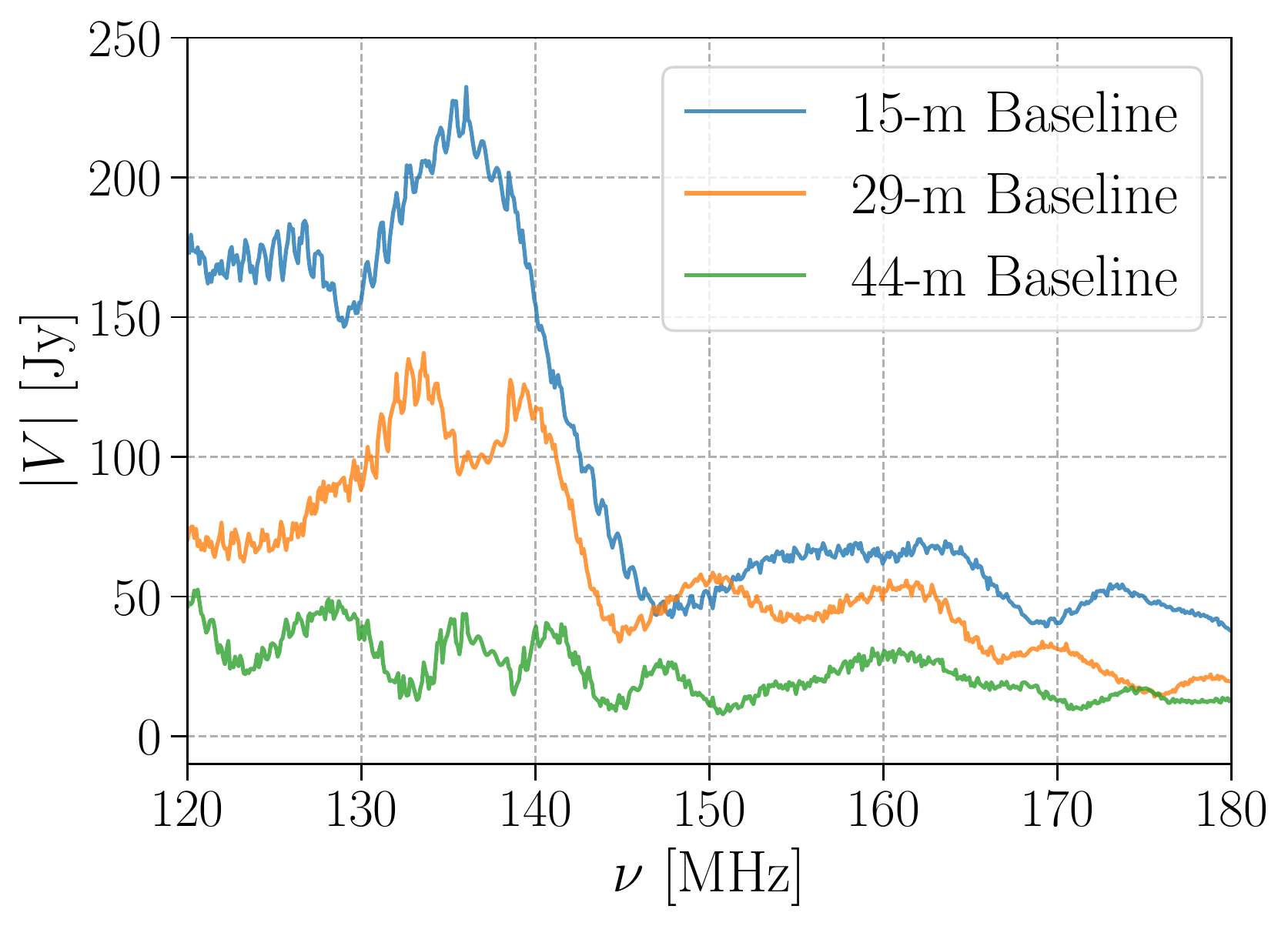}
\caption{Cross correlation visibility amplitudes in frequency space for three East-West oriented baselines increasing in length from 15 meters up to 44 meters at an LST of $\sim6$ hours.
In addition to a broad-scale ripple that decreases in spectral scale with increasing baseline length (most apparent at lower frequencies), we can also see a fast ripple at roughly a 1 MHz scale in all baselines that is likely due to a cross coupling systematic.}
\end{figure}

\begin{figure*}[t!]
\centering
\label{fig:hera_cross_corr}
\includegraphics[scale=0.5]{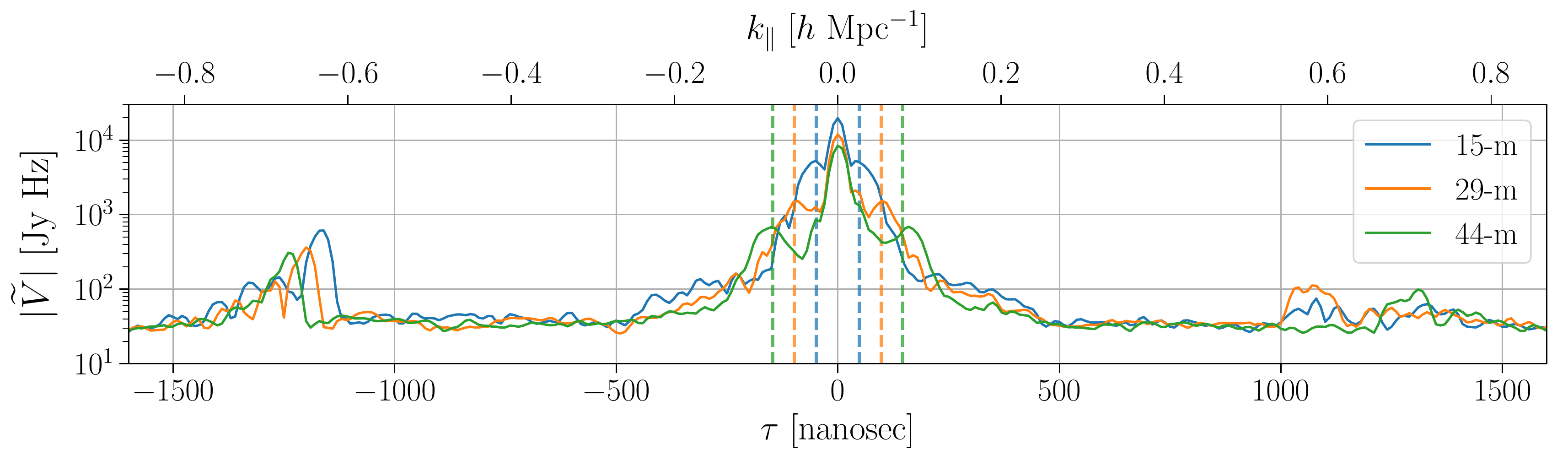}
\caption{HERA cross correlation visibilities averaged in amplitude across LST for three East-West baselines of increasing length: 15 meters, 29 meters and 44 meters (blue, orange and green, respectively).
The dashed vertical lines represent the geometric delay of the horizon for each baseline, within which foreground emission is nominally bounded.
We see spikes in amplitude at the geometric horizon (``low-delay spikes'') and also at higher delays of $|\tau| > 700$ ns (``high-delay spikes'').
The low-delay spikes are thought to be either a pitchfork-effect as predicted by \citet{Thyagarajan2015a} or antenna cross coupling.
Evidence suggests the high-delay features to be some kind of cross coupling systematic.}
\end{figure*}

In a similar fashion as before, we perform a delay-space deconvolution to fill-in missing data due to RFI flags and suppress its sidelobes in the delay domain.
We allow the deconvolution to set model components out to $|\tau| < 1600$ ns, and iterate down to 5$\times$ the noise floor of the visibilities.
\autoref{fig:cross_corr_freq_spectra} shows visibility spectra from the three baselines of interest after deconvolution.
We can clearly see a fast ripple on all baselines with a spectral scale of roughly 1 MHz.
We also see larger scale ripples (particularly at the lower half of the band) that decrease in spectral scale with increasing baseline length.
As we will see below, the former is likely a combination of a cross coupling and reflection systematic, while the latter may also be a form of cross coupling systematic.

Next we window the visibilities from 120 -- 180 MHz with a Blackman-Harris function \citep{Blackman1958} to limit spectral leakage, and then Fourier transform the visibilities to delay space. 
At the moment we are only interested in diagnosing systematics, so we do not square the Fourier amplitudes as we would in forming power spectra, meaning the visibilities are in units of Jansky Hz.
\autoref{fig:hera_cross_corr} shows the result for the 15-meter baseline (blue), 29-meter baseline (orange) and 44-meter baseline (green).
Also plotted as dashed vertical lines are the geometric horizons for each baseline.
The nearly-symmetric peaks at each baseline's geometric horizon could be due to the ``pitchfork'' effect predicted to exist for wide-field radio interferometers \citep{Thyagarajan2015a}.
The pitchfork effect is not a systematic in the context of this work: it is a natural phenomenon from diffuse foregrounds, and is explained as the boosting of measured diffuse sky power near the horizon, where sky signal shows up in the visibilities with delays of the baseline's geometric horizon.
While HERA has a more compact primary beam compared to other low-frequency \tocm experiments (e.g. MWA, PAPER), the pitchfork effect was nonetheless predicted to exist from simulations of the HERA dish and feed \citep{Thyagarajan2016}.
However, these features could also be due to sky emission reflecting off the feed of one antenna and entering the feed of a neighboring antenna (i.e. feed-to-feed reflections), which is a form of antenna cross-coupling that we would also expect to appear at the delay of each baseline's geometric horizon.
While both are expected to produce power at a baseline's geometric horizon, both are also expected to be slowly time-variable, meaning they will occupy similar modes in the delay \& fringe-rate Fourier domains.\footnote{Cross coupling produces slowly time-variable signals in the visibility because it inserts a copy of the auto-correlation, which is slowly time variable). The pitchfork mechanism is a mimicking of the auto-correlation at declinations near the horizon, thus we expect it to have a slow time variability like the auto-correlation.}

\begin{figure*}
\centering
\label{fig:cross_corr_sim_comparison}
\includegraphics[scale=0.45]{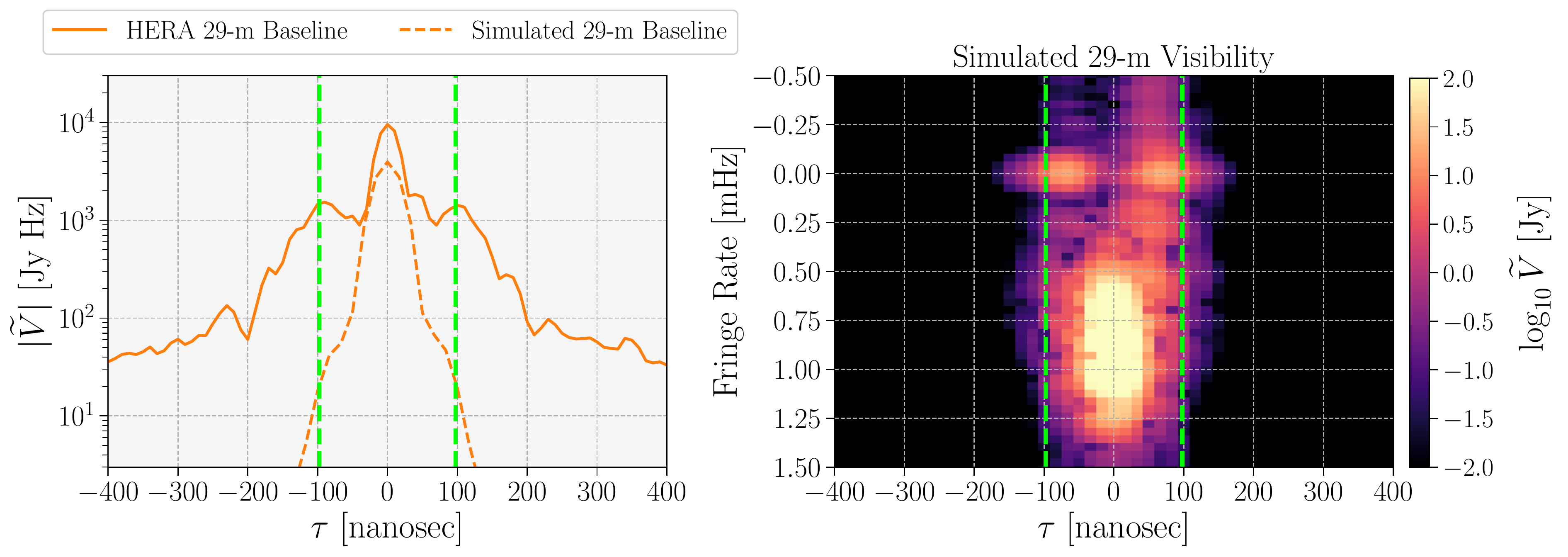}
\caption{Comparison of HERA data with a simulated foreground visibility using the diffuse GSM sky for a 29-meter East-West baseline.
{\bfseries Left:} Averaged HERA cross correlation visibility amplitude in delay space (solid) with an equivalent data product from a simulated foreground visibility with matching LST range (dashed). The geometric baseline horizon is shown at $\sim100$ ns (dashed green). While we see some evidence for a slight pitchfork-like structure in the simulated visibility, it is significantly weaker than the power bumps at equivalent delays in the real data.
{\bfseries Right:} The simulated visibility transformed to fringe-rate and delay space, with the geometric baseline horizon over-plotted (dashed green). We can more clearly see the existence of the pitchfork effect in this plot, which is centered at-$f=0$ mHz and extends out to the natural geometric horizon and quickly falls off after.}
\end{figure*}

\begin{figure*}
\centering
\label{fig:hera_cross_corr_frate_dly}
\includegraphics[scale=0.5]{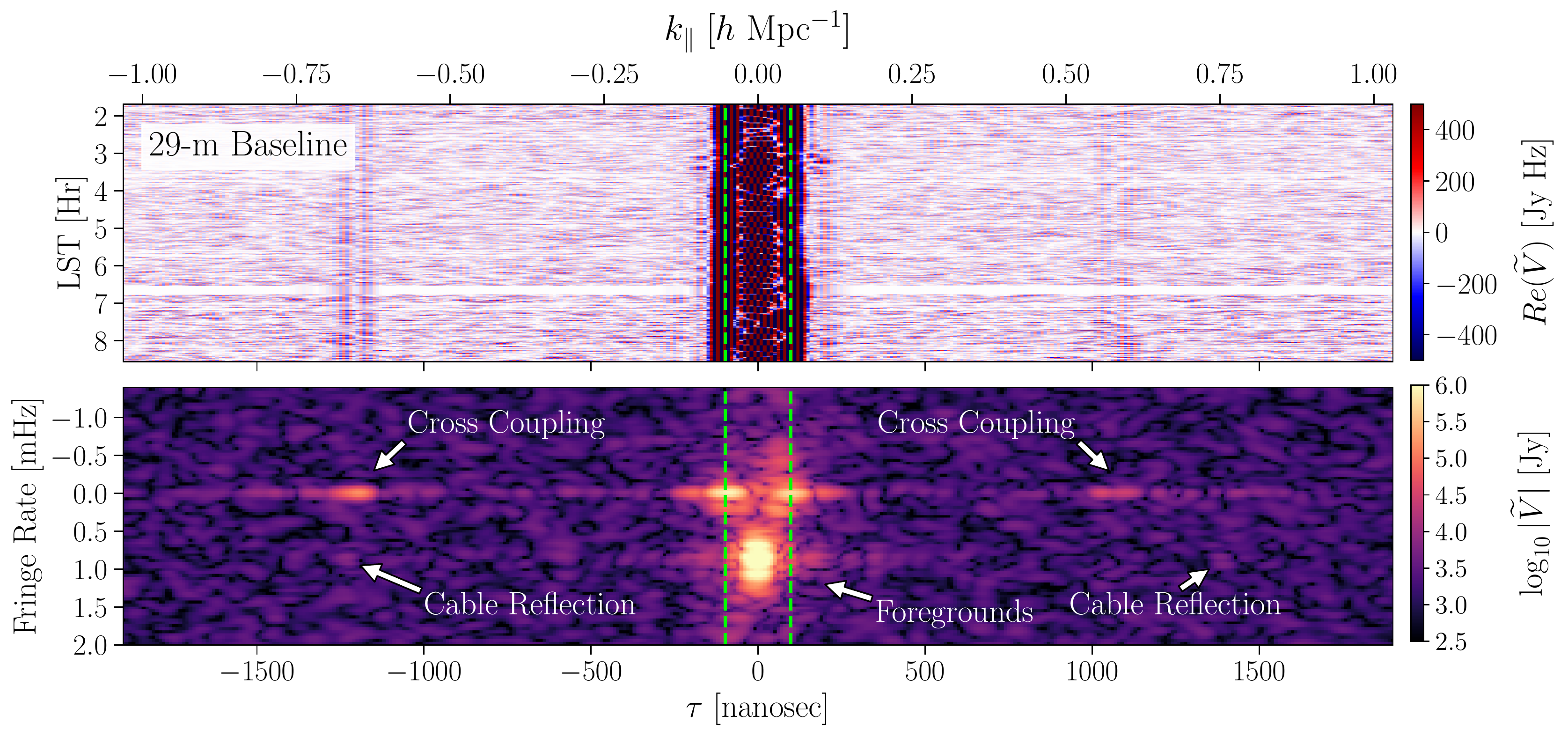}
\caption{A HERA cross correlation visibility showing foregrounds, cable reflections and cross coupling systematics.
{\bfseries Top:} Real component of the visibility in time and delay space, showing foreground power falling within the geometric horizon (green dashed). Notice that power well within the horizon fringes quickly as a function of time, while power near the geometric horizon shows much slower time variability and has spillover to outside the baseline's horizon.
{\bfseries Bottom:} Visibility amplitude in fringe-rate and delay space. Here, we can see the slowly time variable systematics confined to $f\sim0$ mHz fringe-rate modes, while foreground power is boosted to positive fringe rates. In addition, although not visible in the top plot, we can see the cable reflection just barely visible from the background noise, which appears at positive fringe-rates because it is merely a copy of the intrinsic foreground signal.}
\end{figure*}

In \autoref{fig:cross_corr_sim_comparison} we compare the data against a simulated diffuse foreground visibility from \citet{Kern2019a}, which uses the Global Sky Model \citep{Oliveira2008} as the foreground model and a simulated direction-dependent primary beam response for HERA \citep{Fagnoni2019}.
While we do see evidence for a slight pitchfork effect in the simulated data at the geometric delay, its amplitude is considerably weaker than what is observed in the data.
There is also some total power missing from the $\tau=0$ mode, which is likely due to our exclusion of point sources in the simulation.
The simulated pitchfork can be seen more clearly when transforming the simulated visibility into fringe-rate and delay space (right of \autoref{fig:cross_corr_sim_comparison}), where indeed we see the pitchfork occupying $f\sim0$ mHz modes as expected.
As noted, this result is at odds with previous work predicting a strong pitchfork effect in HERA data \citep{Thyagarajan2016}, which used a different model for the HERA primary beam.
This comparison needs further study before we can unequivocally state that the excess power at the geometric horizon is feed-to-feed cross coupling in nature: the simulated pitchfork is highly dependent on the adopted primary beam response at the horizon, which is typically the least accurate aspect of the simulated primary beam response and is also hard to characterize empirically.
A more rigorous analysis using a combination of empirical primary beam constraints as well as a suite of primary beam simulations is needed to better understand this effect in HERA data.

We also see evidence in \autoref{fig:hera_cross_corr} for non-negligible amounts of spillover of foreground emission (or supra-horizon emission) beyond the baseline's geometric horizon, which has also been observed by other \tocm experiments \citep[e.g.][]{Pober2013a, Beardsley2016}.
Supra-horizon emission can come naturally from intrinsic spectral structure of the foregrounds.
It can also be created by chromaticity of the instrumental gain that pushes out structure inherently contained within the geometric horizon, or from low-level artifacts in the data which have a similar effect \citep{Offringa2019}.
As noted above, the antenna-based gains we apply to the data are simplified to a single flux scaling and a single delay, meaning a large part of the observed supra-horizon emission is likely due to uncalibrated instrumental gain terms, which we do not explore in this work.
For a foreground-avoidance approach to estimating the \tocm power spectrum, the presence of supra-horizon emission is highly concerning because it limits our ability to measure the low $k$ modes that in theory probe the EoR at the highest signal-to-noise ratio.
The upside is if supra-horizon emission is slowly time-variable (as are both the pitchfork effect and antenna cross coupling systematics), then regardless of its origin we can mitigate it by filtering it off in Fourier space.
Indeed, this is exactly the principle that cross-coupling subtraction algorithms are founded upon.

Another striking feature in \autoref{fig:hera_cross_corr} is the large amount of excess power above the noise floor at high delay ($|\tau| > 700$ ns).
These features, which we refer to as the ``high-delay'' spikes, exhibit some very peculiar behavior.
First, these features seem to be highly baseline-dependent: the three baselines shown in this section are all tied to antenna 11, yet their structures do not seem to be significantly correlated between the baselines.
Second, their profile as a function of delay does not show isolated, individual peaks as one might expect from one or a few feed-to-feed reflections, but rather shows a wide range of delays corrupted by excess power.
Third, while the structures show up roughly near the delays where we would expect reflections from the 150-m cable to appear, they also show up at delays significantly smaller, enough to necessitate a considerably shorter cable length than 150 meters, which is unlikely.
The high-delay spikes exhibit show slow time-variability with their power centered at $f=0$ mHz, as we would expect from a cross coupling systematic.
\autoref{fig:hera_cross_corr_frate_dly} shows the cross-correlation visibility from the 29-meter baseline in time \& delay space (top) as well as in fringe-rate \& delay space (bottom), where recall the latter is merely the Fourier transform of the former across time.
We can clearly see that the high-delay structures are slowly variable, both by their slow movement as a function of time in the top plot, but also by the fact that their power is centered at $f=0$ mHz in the bottom plot.
This is in contrast to the foreground power centered at $\tau=0$ ns, which oscillates rapidly as a function of time and is therefore boosted to positive fringe-rates, with the exception of the power at the baseline's geometric horizon (dashed green), which, like the systematics at high delay, exhibits slow time variability centered at $f=0$ mHz.

\begin{figure*}[t!]
\centering
\label{fig:hera_svd_modes}
\includegraphics[scale=0.5]{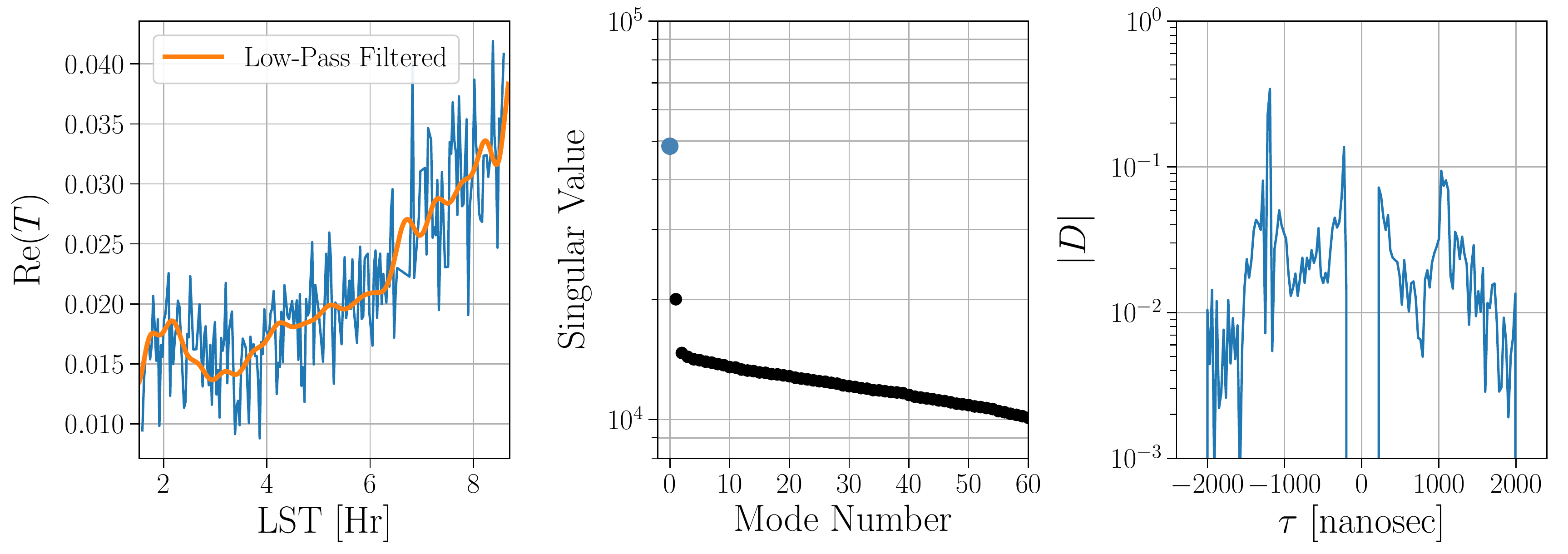}
\caption{Singular value decomposition of the 29-m East-West baseline visibility from \autoref{fig:hera_cross_corr_frate_dly}.
{\bfseries Left:} The first $\mathbf{T}$ eigenvector across time showing its raw form (blue) and its low-pass filtered form (orange), having filtering out modes with $f>0.46$ mHz with a Gaussian Process model \citep{Kern2019a}.
{\bfseries Center:} The first sixty singular values, showing that most of the variance in the systematic-prone regions can be described with a handful of modes before a noise plateau is reached.
{\bfseries Right:} The first $\mathbf{D}$ eigenvector across delay, showing it picking up on the slowly variable structure at large delays ($|\tau|\sim1200$ ns) and also some structure near the baseline horizon ($|\tau|\sim200$ ns).}
\end{figure*}

What we cannot see by looking at the visibility in time \& delay but can barely begin to discern when we transform to the fringe-rate domain are the cable reflections at $|\tau|\sim1300$ ns.
As we saw in \autoref{fig:autoamp_avg}, the measured reflection amplitudes are roughly $3\times10^{-3}$ times the peak power in the visibility.
Because the high-delay spikes at $f=0$ mHz also show up at similar delays and are stronger in amplitude, we cannot see the cable reflections in \autoref{fig:hera_cross_corr} or in the top panel of \autoref{fig:hera_cross_corr_frate_dly} buried under the other systematics.
Reflections have the same time-structure as the unreflected signal, so by transforming to fringe-rate space we can isolate them from the slowly time variable systematics, and indeed we can just barely seem them above the noise floor of the cross correlation visibilities at roughly $3\times10^{-3}$ times the main foreground power as expected.
\autoref{fig:hera_cross_corr_frate_dly} also shows evidence for the supra-horizon emission having two distinct components: one that is has fast time variability like foregrounds from the main-lobe of the primary beam, and another that is slowly fringing like a cross coupling systematic or a pitchfork effect, and both extend considerably beyond the baseline's geometric horizon.

\begin{figure*}
\centering
\label{fig:hera_cross_corr_sub}
\includegraphics[scale=0.5]{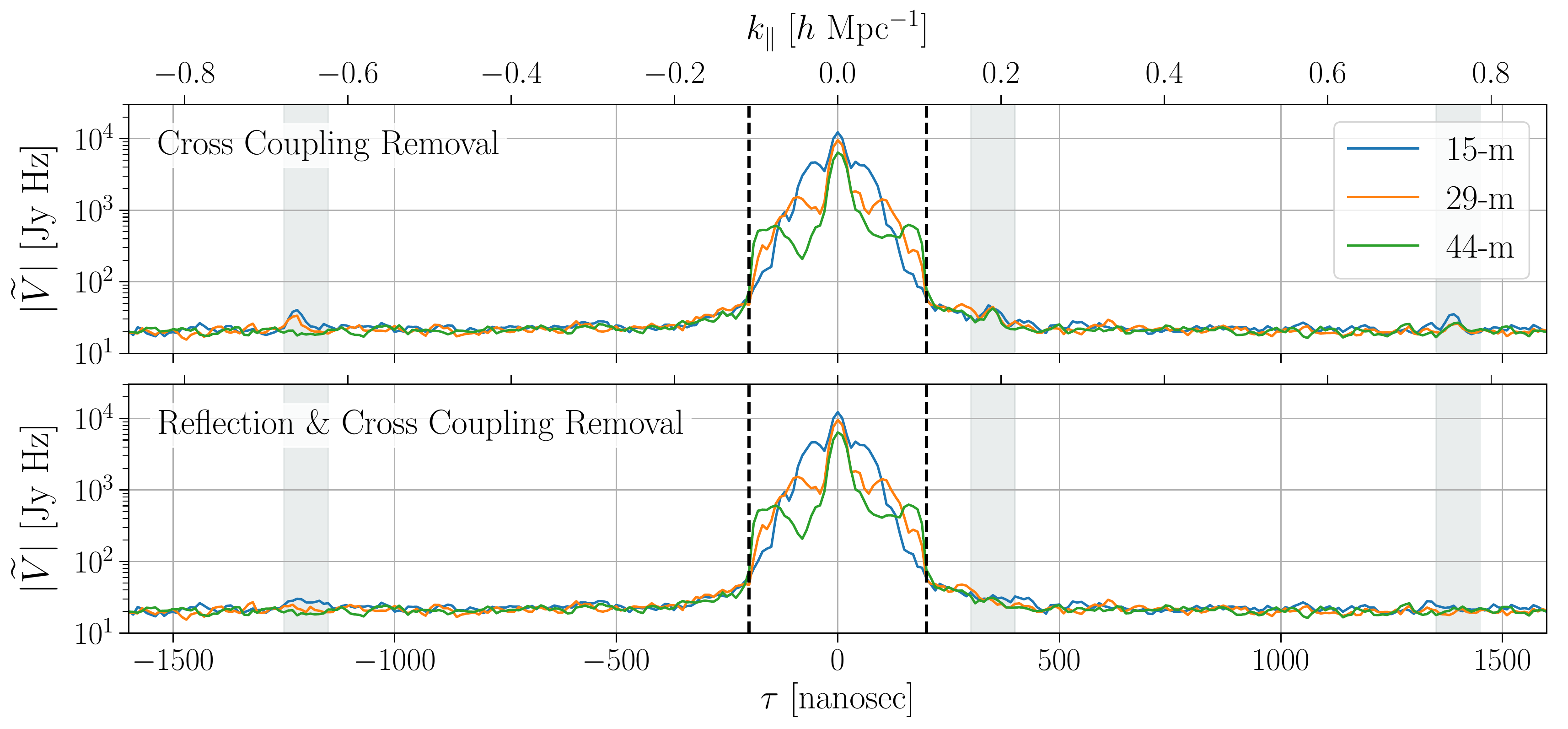}
\caption{HERA cross correlation visibilities from \autoref{fig:hera_cross_corr} after cross coupling subtraction but before reflection calibration (solid) and after both cross coupling subtraction and reflection calibration (dashed).
The black-dashed line represents the lower delay boundary of the cross coupling model.
Grey shaded regions indicate expected delays for reflection systematics having inspected the auto-correlations for peaks.
Joint systematic suppression yields cross correlations visibly free of systematics at the level of the per-baseline noise floor.}
\end{figure*}

\begin{figure*}
\centering
\label{fig:hera_cross_corr_frate_dly_sub}
\includegraphics[scale=0.5]{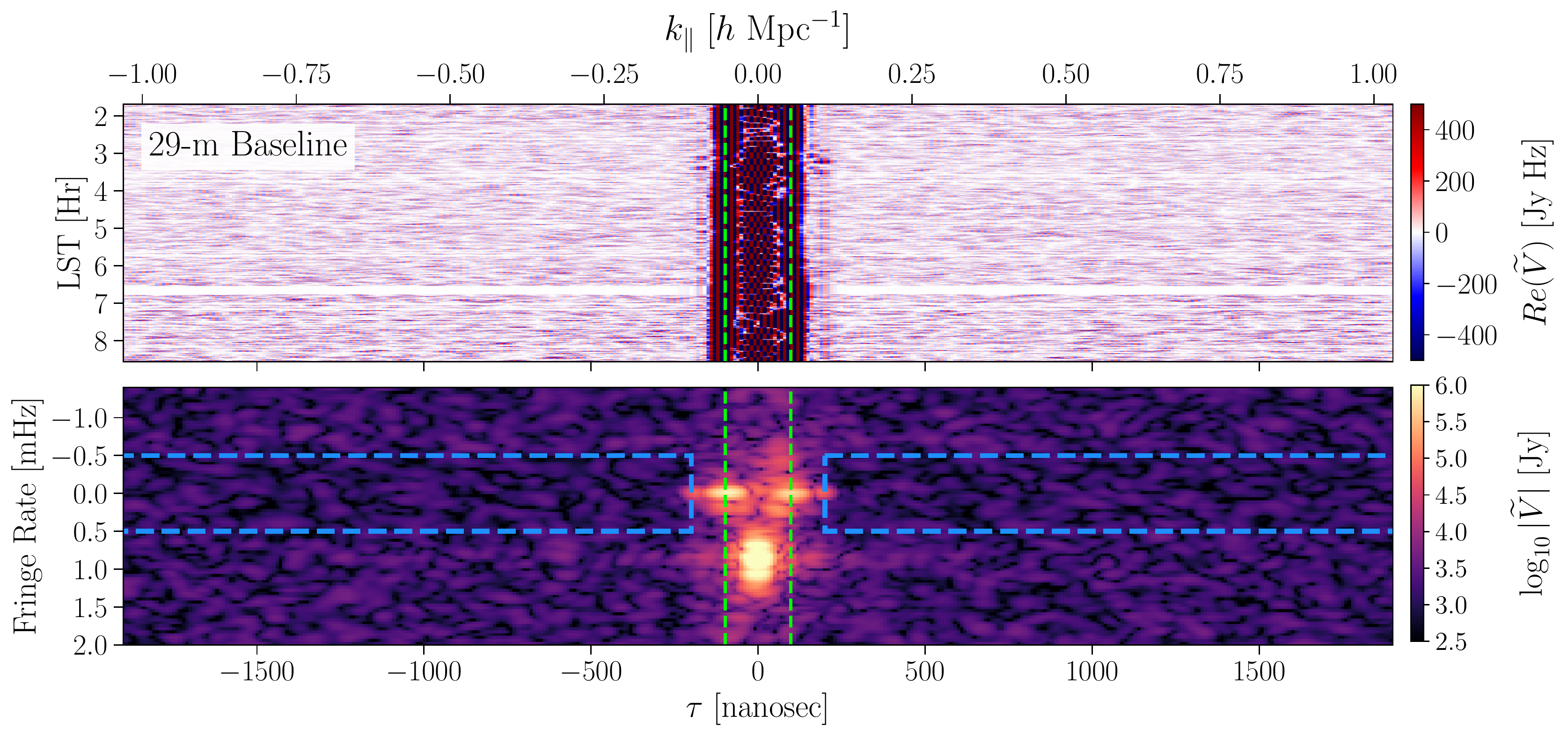}
\caption{Same 29-m visibility in fringe-rate and delay space as shown in \autoref{fig:hera_cross_corr_frate_dly} but now with reflection and cross coupling systematics removed. The blue-dashed region shows where the cross coupling algorithm modeled and removed systematics, and the green-dashed line marks the baseline's geometric horizon.}
\end{figure*}

Currently, there is not a single physical model for the origin of the high-delay spikes that can explain all of its behavior observed in the data.
In \autoref{sec:cross_coupling_models}, we explore some simple physical models for the systematic and show that we can tentatively rule them out; however, further work is needed to more fully understand their origin.
Nonetheless, their temporal behavior is suggestive of some kind of antenna cross coupling that occurs at some point along the signal chain.
At present, what we can say with certainty is that their time-dependence is highly inconsistent with an EoR signal, and as such we can suppress it by filtering the data in fringe-rate space before forming power spectra (see \citet{Kern2019a} for details on why this is inconsistent with an EoR signal).

With that in mind, \autoref{fig:hera_svd_modes} shows the result of running an SVD-based cross coupling model \citep{Kern2019a} on the 29-meter baseline data, which decomposes the matrix shown in the top panel of \autoref{fig:hera_cross_corr_frate_dly} into orthogonal time eigenmodes ($\mathbf{T}$), orthogonal delay eigenmodes ($\mathbf{D}$) and their singular values ($\mathbf{S}$).
Before taking the SVD we apply a bandstop window on the data matrix that assigns zero weight to all delay modes outside of the range $200 < |\tau| < 2000$ ns, which was chosen to encompass most of the observed cross-coupling systematics and to reject the foregrounds at very low delays.
The left panel plots the first $\mathbf{T}$ eigenmode across time, showing the raw eigenmode (blue) and the eigenmode after low-pass filtering it out to $f_{\rm max} = 0.46$ mHz (orange).
We use the Gaussian Process-based filter explored in \citet{Kern2019a} to low-pass filter these time-modes.
The center panel shows the first 60 singular values, giving us a sense for how much the information content is isolated into the first few eigenmodes.
We find that most of the structure can be described with only a handful of modes before reaching a plateau.
In forming the systematic model we keep the top 30 modes out of $\sim1000$ and truncate the rest.
Lastly, the right panel shows the first $\mathbf{D}$ eigenmode across delay, showing it picking up the high-delay cross coupling systematic and some of the supra-horizon emission at low delay.
In addition to picking up on the systematic, the SVD will pick up on the noise of the data as well.
However, because we keep only a small fraction of the eigenmodes and additionally smooth them across time, we do not suspect that we are subtracting a significant component of the noise in the process of systematic removal.

We repeat this for the other baselines at hand, low-pass filtering the $\mathbf{T}$ basis vectors from the 15-meter and 44-meter baselines with $f_{\rm max} = 0.14$ and $0.83$ mHz respectively, using a Gaussian-process-based smoothing for the low-pass filter.
See Table 1 and Appendix B of \citet{Kern2019a} for more details on this process.
\autoref{fig:hera_cross_corr_sub} shows the baselines in \autoref{fig:hera_cross_corr} after cross-coupling subtraction, with the vertical dashed line showing the minimum delay of the cross coupling model at $\tau=200$ ns.
The top panel shows only cross coupling subtraction, where we see significant suppression of the high-delay spikes and the outer edge of the low-delay spikes.
As expected, after subtracting the strong cross coupling terms at high-delay we are left with the appearance of localized bumps that mark the cable reflections (marked in grey bands), which recall were not subtracted out with the cross coupling because they occupy fringe-rate modes that were filtered out of the systematic model in the process of smoothing.
The bottom panel shows the data after applying reflection calibration from \autoref{sec:ref_cal} and cross coupling subtraction, showing that the data is now consistent with a scale-independent thermal noise floor for all delays outside $|\tau| > 500$ ns.

There is, however, still a slight slope in the data at intermediate delays of $200 < |\tau| < 500$ ns, which is part of the supra-horizon emission we observed earlier.
To ensure that this tail is not coming from the cross-coupling component that we attempted to filter out, we can plot the systematic-subtracted data in fringe-rate \& delay space, which is shown in \autoref{fig:hera_cross_corr_frate_dly_sub} with the blue-dashed region showing the region of Fourier space where cross coupling subtraction was performed.
\autoref{fig:hera_cross_corr_frate_dly_sub} confirms that the excess signal between $200 < |\tau| < 500$ ns observed in \autoref{fig:hera_cross_corr_sub} does not come from modes that should have been subtracted in the process of cross coupling removal, and originates from the second supra-horizon component at higher fringe-rates.
As discussed above, this supra-horizon emission can come from uncalibrated bandpass terms or from low-level artifacts in the data, which push foregrounds out in delay that were intrinsically contained within the geometric horizon.
These effects can be somewhat mitigated with better bandpass calibration and data flagging, but are still active areas of research in the literature.
Additionally, the slight overlap of low fringe-rate power inside the dashed region at $|\tau|=200$ ns is produced by a windowing function applied to the data before taking its Fourier transform.

Signal loss is a principal concern when applying any baseline-dependent operation to the data, as we have done with cross coupling subtraction.
In \citet{Kern2019a} we vet our cross coupling modeling algorithms for EoR signal loss against numerical visibility simulations of the HERA Phase I system.
We show that by low-pass filtering the systematic model along time (\autoref{fig:hera_svd_modes}), we can harden our systematic model against EoR signal loss to an almost arbitrary level.
In our case, we chose the fringe-rate bounds above by adopting a signal loss tolerance of 1\% in EoR power, which is below the expected measurement error of the full HERA array.
We refer the reader to our analysis and discussion in that paper for more details on signal loss quantification in the context of cross coupling removal.

\section{Power Spectrum Estimation}
\label{sec:power_spectra}
Now that we've demonstrated that we can suppress reflection and cross coupling systematics for a few baselines down to their individual noise floors, we would like to prove that we can similarly do this for baselines across the entire array, and confirm that these systematics are a non-limiting factor in the power spectrum even after redundant baseline averaging.
We will focus on the same three baseline orientations (14-m, 29-m and 44-m East West baselines), but now look at all baselines within the array that fall within each baseline group.

\subsection{Delay Spectra}
To estimate the three-dimensional \tocm power spectrum, $P_{21}(\mathbf{k})$, we use the delay spectrum estimator \citep{Parsons2012b, Liu2014a, Parsons2014}.
The delay spectrum is a per-baseline, visibility-based power spectrum estimator that relies on the Fourier transform of the visibility across frequency into the delay ($\tau$) domain,
\begin{align}
\label{eq:delay_transform}
\widetilde{V}(\mathbf{u}, \tau) = \int d\nu\ e^{2\pi i\nu\tau} V(\mathbf{u}, \nu),
\end{align}
where $\mathbf{u} = \mathbf{b} / \lambda$ is the baseline vector divided by the observing wavelength.
The ``delay transform'' of the visibility is not a direct measurement of the line-of-sight cosmological $\kpara$ mode, due to an interferometer's inherent chromaticity \citep{Morales2012}.
Approximating it as such is known as the ``delay approximation,'' which was shown to be a good approximation for short baselines and is one of the motivating factors behind HERA's compact design \citep{Parsons2012b, Dillon2016}.
We refer the reader to \citet{Morales2019} for a broader discussion on various \tocm power spectrum estimators.
The delay spectrum approximation of the \tocm power spectrum is then the square of the delay-transformed visibility with the appropriate scaling factors,
\begin{align}
\label{eq:delay_spectrum}
\widehat{P}_{21}(\kperp, \kpara) \approx |\widetilde{V}(\mathbf{u}, \tau)|^2\frac{X^2Y}{\Omega_{pp}B_{p}}\left(\frac{c^2}{2k_{B}\bar{\nu}^2}\right)^2
\end{align}
where $X$ and $Y$ are redshift-dependent scalings converting sky angles and frequencies to cosmological length scales, $\Omega_{pp}$ is the sky-integral of the squared antenna primary beam response, $\bar{\nu}$ is the delay transform center frequency and $B_{p}$ is the delay transform bandwidth, as defined in Appendix B of \citet{Parsons2014}.
The factors relating the $\mathbf{u}$ and $\tau$ Fourier domains inherent to the telescope to the cosmological Fourier domains of $\kperp$ and $\kpara$ are
\begin{align}
\label{eq:cosmo_scalings}
\kpara &= \frac{2\pi}{X}\tau \nonumber \\
\kperp &= \frac{2\pi}{Y}\frac{b}{\lambda}
\end{align}
where $X = c(1+z)^2\nu_{21}^{-1}H(z)^{-1}$, $Y = D(z)$, $\nu_{21}=1.420$ GHz, $H(z)$ is the Hubble parameter, $D(z)$ is the transverse comoving distance, $b$ is the baseline length and $\lambda$ is the observing wavelength \citep{Parsons2012a, Liu2014a}.

Cross multiplying a visibility with itself in \autoref{eq:delay_spectrum} to form a delay spectrum will result in an overall bias in power due to the noise present in the data.
To avoid this, we take visibility spectra adjacent to each other in LST separated by 10.7 seconds and apply a phasing term to align their phase centers before cross multiplication \citep{Pober2013b}.
This means the two visibilities to leading order measure the same cosmological mode on the sky but have uncorrelated noise realizations, such that they do not produce a noise bias upon cross correlation.


%

\begin{figure}
\centering
\label{fig:hera_tsys}
\includegraphics[scale=0.55]{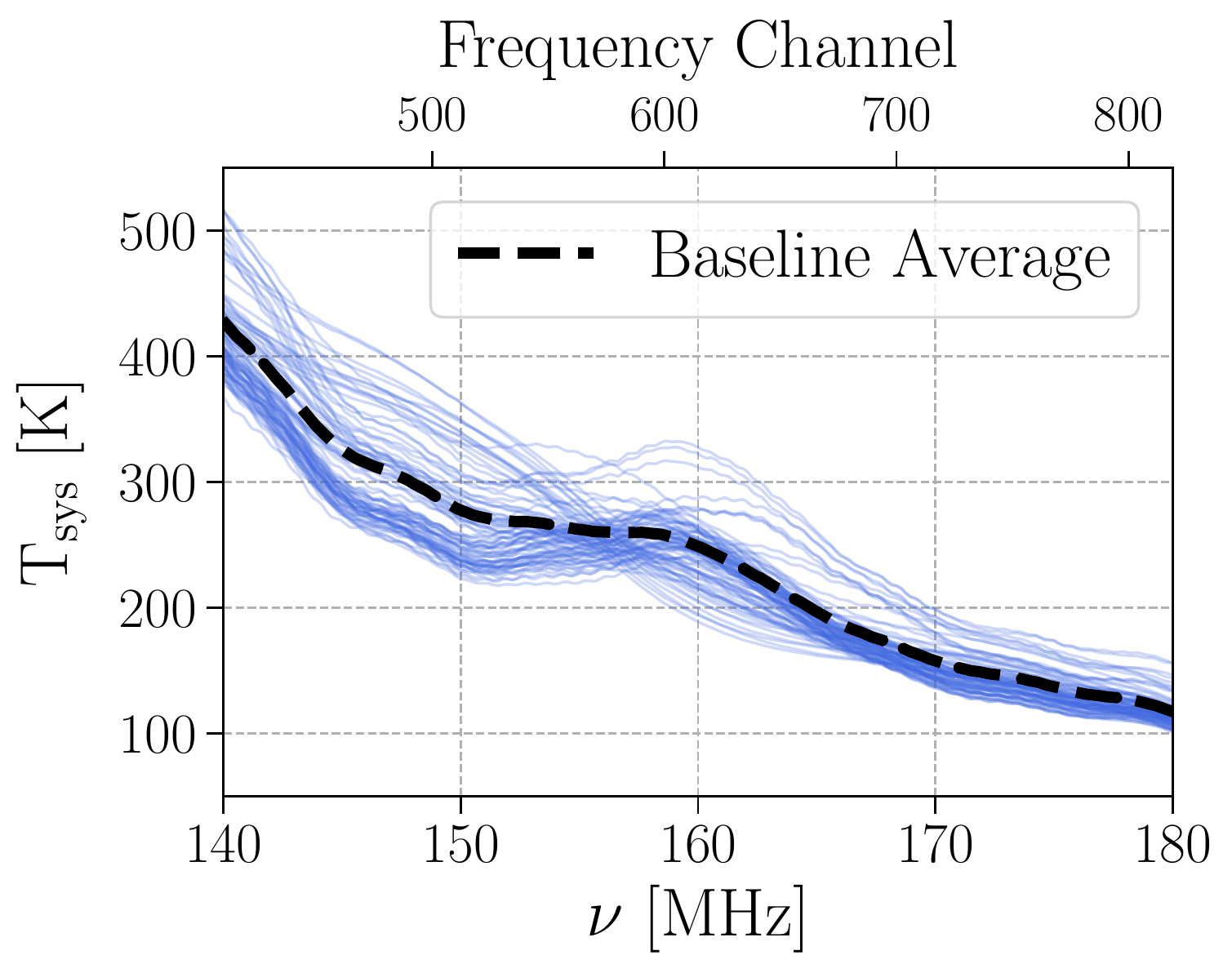}
\caption{System temperature curves for all baselines used in the power spectral analysis (colored points), and their average (black dashed). Delay spectra presented in this section are formed between channels 450 and 650 (144 -- 163 MHz) with an effective system temperature of $\sim270$ K.}
\end{figure}

\begin{figure*}
\centering
\label{fig:pspec_waterfall}
\includegraphics[scale=0.5]{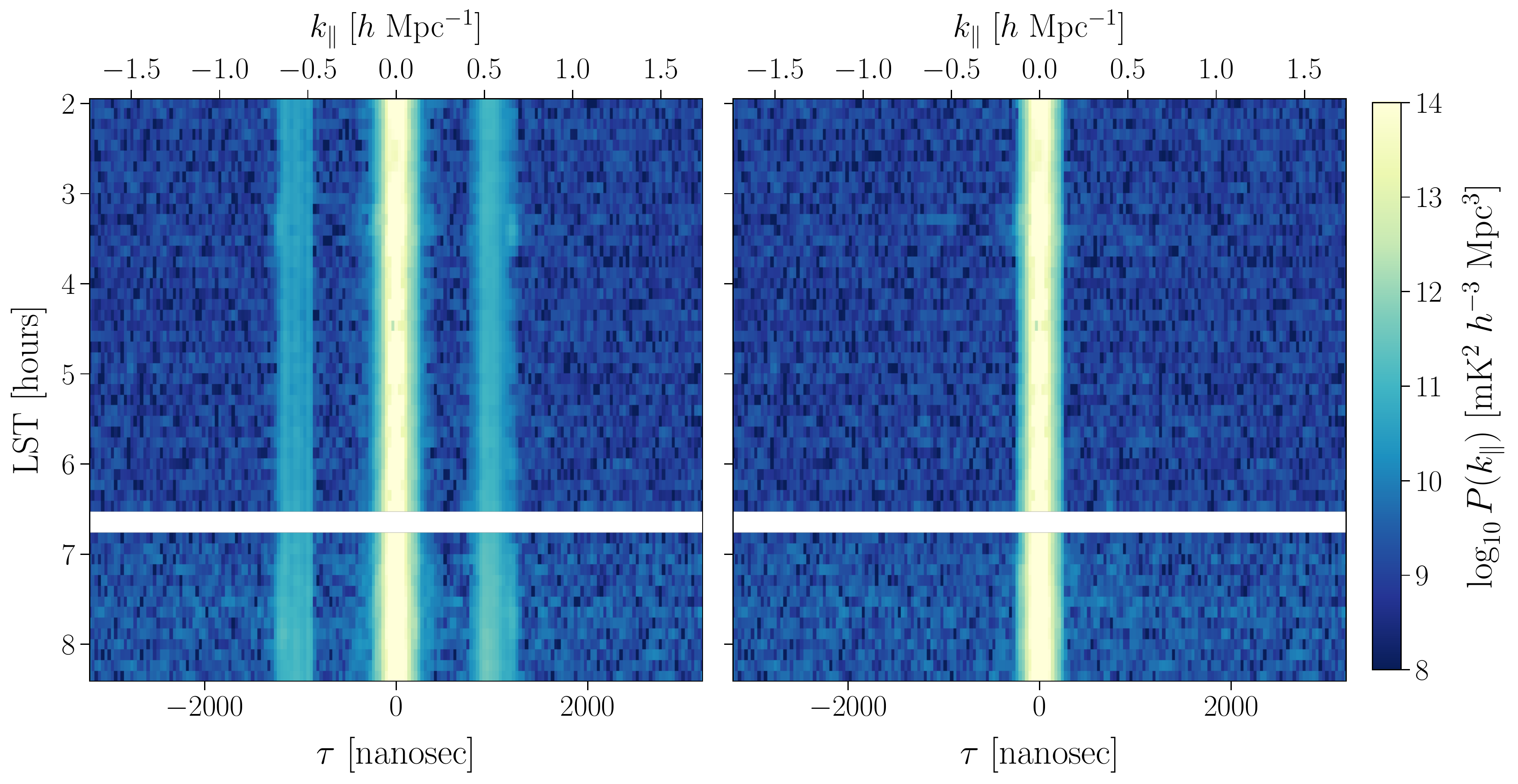}
\caption{An averaged power spectrum waterfall of the East-West 15-m group showing the absolute value of the real component of the power spectra, having first incoherently averaged 35 separate baseline-pairs in the group. We plot the data with systematics in (left) and with systematics removed (right).}
\end{figure*} 

Thermal noise in interferometric visibilities is mean-zero, Gaussian distributed, and is statistically uncorrelated on all time and frequency scales; however, it generally is non-stationary, and will have an amplitude dependence as a function of LST and frequency.
A signal chain's \emph{system temperature} is proportional to the total amount of noise power received by the analogue system, and is the sum of the sky noise and receiver noise,
\begin{align}
T_{\rm sys}(\nu, t) = T_{\rm sky}(\nu, t) + T_{\rm rcvr}(\nu, t)\ {\rm [K]}.
\end{align}
In practice, antenna signal chains will have variable system temperatures due to different angular primary beam responses and different receiver properties.
A visibility-based system temperature can therefore be estimated, which is the system temperature as measured by a particular baseline.
This can be estimated by taking differences of adjacent pixels in time and frequency and relating its RMS to a system temperature via the radiometer equation,
\begin{align}
\sigma_{\rm rms}^{ij} = \frac{2k_b\nu^2}{c^2\Omega_p} \frac{T_{\rm sys}^{ij}}{\sqrt{\Delta\nu\Delta t}},
\end{align}
where $\sigma_{\rm rms}^{ij}$ is the RMS of the visibility between antennas $i$ and $j$ in Jansky, $k_b$ is the Boltzmann constant, $\nu$ is the average observing frequency, $\Omega_p$ is the angular integral of the peak-normalized primary beam response in steradians, $\Delta\nu$ is the correlator channel width in Hz and $\Delta t$ is the correlator integration time in seconds \citep{Thompson2017}.
Another estimate of the noise comes directly from the auto-correlation visibility, which itself is a measurement of the total power received by a particular antenna.
For a cross-correlation visibility between antenna $i$ and $j$, we can estimate the baseline's system temperature as
\begin{align}
\sqrt{V_{ii}V_{jj}} = \frac{2k_b\nu^2}{c^2\Omega_p}T_{\rm sys}^{ij},
\end{align}
where $V_{ii}$ is the auto-correlation visibility of antenna $i$.
While both methods are comparable, we defer to using the auto-correlations, which in practice generally lead to more stable and cleaner noise models.

\autoref{fig:hera_tsys} shows system temperature estimates for each baseline participating in the analysis (blue) and the averaged system temperature, which is each baseline's system temperature averaged in quadrature.
Again, because we have not corrected for the bandpass structure of the gains, the large-scale fluctuations in \autoref{fig:hera_tsys} are not unexpected, and would be smoothed-out after solving for and applying the appropriate instrumental gains.
The presence of such structure in the noise curves does not change the fundamental results of this section.
Power spectra presented in this section are formed between channels 450 -- 650 (144 -- 163 MHz) with an effective system temperature of $\sim270$ K.

\begin{figure*}
\centering
\label{fig:hera_avg_pspec}
\includegraphics[scale=0.5]{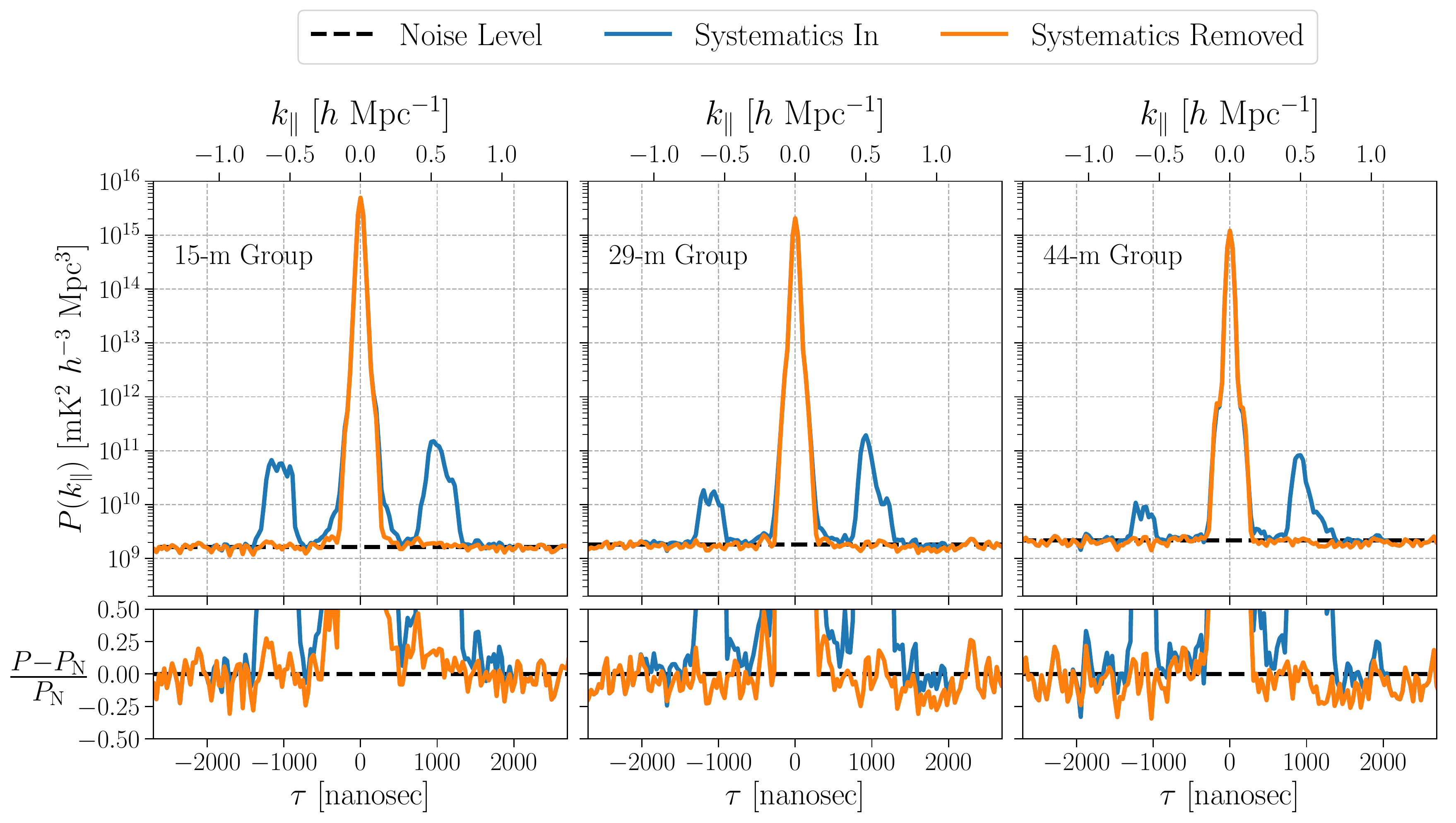}
\caption{Delay spectra for three unique baseline lengths oriented along the East-West axis without systematic removal (blue) and with systematic removal (orange). The power spectra are formed directly from the visibilities for each baseline in the array, are incoherently averaged within each redundant group, and then their absolute value is averaged across the remaining bins in LST.
We see suppression of high delay systematics down to the integrated noise floor, and get some suppression of supra-horizon power at low delay.}
\end{figure*}

With an understanding of the noise properties of our data, we can compute a theoretical estimate of the noise power spectrum, $P_{\rm N}$, which is equivalent to the root-mean square (RMS) of the power spectrum if the only component in the data were noise.
This is one way to measure the uncertainty on the estimated power spectra, but also represents the theoretical amplitude of the power spectra in the limit that they are noise dominated (as opposed to signal or systematic dominated).
This is given in \citet{Cheng2018} as
\begin{align}
\label{eqn:PN}
P_{\rm N} = \frac{X^2Y\Omega_{\rm eff}T_{\rm sys}^2}{t_{\rm int}N_{\rm coherent}\sqrt{2N_{\rm incoherent}}},
\end{align}
where the $X$ and $Y$ scalars are the same as before, $T_{\rm sys}$ is the system temperature in milli-Kelvin, $t_{\rm int}$ is the correlator integration time in seconds, $N_{\rm coherent}$ is the number of sample averages done at the visibility level (i.e. before visibility squaring), and $N_{\rm incoherent}$ is the number of sample averages done at the power spectrum level (i.e. after visibility squaring).
$\Omega_{\rm eff}$ is the effective beam area given by $\Omega_{\rm eff} = \Omega_{p}^2 / \Omega_{pp}$, where $\Omega_{p}$ is the integral of the beam across the sky in steradians, and $\Omega_{pp}$ is the integral of the squared-beam across the sky in steradians \citep{Pober2013b, Parsons2014}.
We calculate $P_N$ for each redundant group using the baseline-averaged system temperature.

The data are natively sampled at a 10.7 second cadence.
Before forming power spectra, we coherently average each visibility across LST for 3.6 minutes (20 samples), applying a fringe-stop in each averaging window to limit sky signal attenuation.
We select a wide spectral window between channels 400 to 700 (139 -- 168 MHz), and apply a Blackman-Harris windowing function before transforming to Fourier space.
Because the cosmological signal undergoes non-negligible evolution within such a bandwidth we would not normally use such a wide bandwidth for setting upper limits, however, we do this to achieve better resolution in delay space for diagnostic purposes.
We then cross-multiply the visibilities and apply the necessary normalization factors as per \autoref{eq:delay_spectrum}.
For simplicity, we only form power spectra by cross multiplying baselines with themselves (at adjacent times), and do not cross correlate different baselines within redundant groups.
Then we average the power spectra within each redundant group (i.e. an incoherent average).
For the 15-m, 29-m and 44-m groups this involves averaging 35, 28, and 20 independent baselines, respectively.

What we are left with is a single complex-valued power spectrum waterfall for each redundant group as a function of LST and delay, consisting of 60 leftover time bins and 200 delay bins.
In \autoref{fig:pspec_waterfall} we show this for the 15-m group with and without systematic removal (right \& left)
In our final step, we take the real component of each power spectrum waterfall and average its absolute value over the remaining time bins.
This is done to make a higher signal-to-noise measurement of the noise floor at the level of the power spectrum waterfall: we could have gained more sensitivity by not taking the absolute value before averaging, but our point here is to make a visually clearer comparison with the known noise level rather than gain increased sensitivity.
\autoref{fig:hera_avg_pspec} shows the power spectra of the data without systematic removal (blue), with systematic removal (orange) and also shows the theoretical noise level given our visibility noise estimates and taking into account the various forms of averaging before and after squaring the visibilities (black dashed).
In this case, the systematic removal includes both cross coupling subtraction and reflection calibration.
We find that we can suppress the observed systematics by roughly two orders of magnitude in power, enabling us to achieve six orders of magnitude in dynamic range with respect to the peak foreground power for  $|k_\parallel| > 0.2\ h$ Mpc$^{-1}$.

The power spectra show generally good agreement with our prediction of the thermal noise floor for delays considerably outside of the foreground wedge.
Although the geometric horizon for these short baselines is on the order of 50 -- 150 ns, the Blackman-Harris windowing function pushes this out by about +100 ns, such that their effective horizon is on the order of 150 -- 250 ns.
However, we can still see some amount of positive power near the transition region, particularly for the 15-meter group.
This could be due to uncalibrated bandpass terms in the data, low-level artifacts in the data missed by RFI flagging, or residual reflection and cross coupling systematics.
More complete gain calibration and deeper integrations will allow us to investigate this at higher SNR levels.

\citet{Gosh2019} also propose methods for subtracting systematics observed in the HERA Phase I instrument using a Gaussian Process based model.
With their model, they find good subtraction of the systematic down to similar dynamic ranges ($10^6$ in power), at the cost of possible signal loss at the $\sim10\%$ level.
Systematics of a similar nature were also observed in the HERA-19 Commissioning array \citep{Kohn2019}.
However, a direct comparison with this work is difficult because the array was re-configured en route to the Phase I configuration.

As a final note, we would like to clarify how we came to the noise level plotted in \autoref{fig:hera_avg_pspec}.
Noise in the interferometric visibility is a complex Gaussian random variable, meaning that when we form power spectra by squaring the visibilities we are left with a noise component that is drawn from a complex normal-product distribution.
A real-valued normal-product distribution can be shown to be described by a modified Bessel function of the second kind of order 0 \citep{Wells1962, Cui2016}.
A complex-valued normal-product random variable is simply the sum of two real-valued normal-product random variables, which means it's probability density function (PDF) is a convolution of the Bessel function with itself and turns out to be a double-sided exponential distribution.
Therefore, after squaring the visibilities, noise in the power spectrum is drawn exponentially.

However, most power spectrum pipelines will average the data after squaring the visibilities (i.e. incoherent averaging), which will re-Gaussianize the data due to the Central Limit Theorem.
Indeed, to create \autoref{fig:pspec_waterfall} we perform a few dozen incoherent averages across redundant baselines after squaring the visibilities, meaning it is fair to assume the noise in our power spectrum is Gaussian-distributed.
However, in order to collapse our data along the LST axis to form \autoref{fig:hera_avg_pspec} we took the absolute value of the real-component of the power spectrum before averaging.
The absolute value operation transforms the noise from a Normally-distributed, mean-zero random variable into a random variable drawn from a half-Normal distribution, which is no longer mean-zero and has an expectation value of $\sigma\sqrt{\frac{2}{\pi}}$.
Recall from \autoref{eqn:PN} that $P_N$ tells us the expected RMS of the real (or imaginary) component of the complex power spectrum due to thermal noise.
Therefore, the act of taking the absolute value of the real-component of the power spectra and averaging across LST means we need to multiply our final $P_N$ estimate by a factor of $\sqrt{2/\pi}$, which is what is actually plotted in \autoref{fig:hera_avg_pspec} as the black-dashed line.

\section{Summary}
In this work we investigate data from HERA Phase 1 for signal chain reflection and antenna cross coupling systematics.
We find cable reflections on the order of $\sim10^{-3}$ in amplitude, and a systematic tail in the auto-correlation visibilities straddling the EoR window at roughly the $10^{-3}-10^{-4}$ level, which is considerably larger than that expected from simulations of the HERA dish and feed.
If not mitigated, the systematic tail observed in the auto-correlations may prevent HERA Phase I from setting competitive upper limits on the EoR, let alone detecting it.
We show that reflection calibration can help to suppress some of these features by about an order of magnitude in the visibility at specific $k_\parallel$ modes.
The presence of the systematic tail in the auto-correlation may be indicative of highly complex cable sub-reflections that will be hard to calibrate out down to EoR levels, even with the methods demonstrated here.

We also inspect the data for antenna cross-coupling systematics and find that they contaminate the data at high delays near the edge of the targeted EoR modes at $k_\parallel\sim0.5\ h$ Mpc$^{-1}$.
We also find evidence for excess emission at each baseline's geometric horizon that is likely due to either 1) a pitchfork effect \citep{Thyagarajan2016} or 2) feed-to-feed mutual coupling.
These features produce non-negligible spillover into the EoR window and thus need to be controlled for foreground avoidance power spectrum approaches.
We investigate three East-West baselines of increasing length (15-m, 29-m \& 44-m) that exhibit particularly strong systematics, and find that we can model and remove both of the contaminating components in the EoR window down to the integrated noise floor of each baseline.

We then form power spectra from three redundant groups for baselines across the entire array.
We show that by combining reflection calibration and cross coupling subtraction on specific baseline orientations, we can suppress all visible systematics for $k_\parallel > 0.2\ h$ Mpc$^{-1}$, down to the integrated noise floor of the array for a single nightly observation: with the exception of a weak supra-horizon tail at low $k$ that merits further investigation through improved bandpass calibration and RFI flagging. 
Instrumental bandpass calibration for HERA Phase I is explored in \citet{Kern2019c} and \citet{Dillon2019}.

This work shows that the immediate systematics seen in HERA Phase I system can be modeled and dealt with down to a dynamic range of $10^{-6}$ in the power spectrum, even with an extremely simple approach to direction-independent, antenna-based calibration.
While this is reassuring, fiducial EoR levels are expected to appear at dynamic ranges of $\sim10^{-10}$ in the power spectrum for low-$k$ modes \citep{Thyagarajan2016}.
Assuming that the systematics studied here can continue to be subtracted to lower noise levels and barring the appearance of other systematics, this work suggests that fully integrated HERA Phase I may have the potential to set competitive upper limits on the \tocm power spectrum.

\

This material is based upon work supported by the National Science Foundation under Grant Nos. 1636646 and 1836019 and institutional support from the HERA collaboration partners.
This research is funded in part by the Gordon and Betty Moore Foundation.
HERA is hosted by the South African Radio Astronomy Observatory, which is a facility of the National Research Foundation, an agency of the Department of Science and Innovation.
A. Lanman and J. C. P. would like to acknowledge NASA Grant 80NSSC18K0389.
A. Liu acknowledges support  from a Natural Sciences and Engineering Research Council of Canada  (NSERC) Discovery Grant and a Discovery Launch Supplement, as well as the Canadian Institute for Advanced Research (CIFAR) Azrieli Global Scholars program.
Parts of this research were supported by the Australian Research Council Centre of Excellence for All Sky Astrophysics in 3 Dimensions (ASTRO 3D), through project number CE170100013.
G. B. acknowledges funding from the INAF PRIN-SKA 2017 project 1.05.01.88.04 (FORECaST), support from the Ministero degli Affari Esteri della Cooperazione Internazionale - Direzione Generale per la Promozione del Sistema Paese Progetto di Grande Rilevanza ZA18GR02 and the National Research Foundation of South Africa (Grant Number 113121) as part of the ISARP RADIOSKY2020 Joint Research Scheme, from the Royal Society and the Newton Fund under grant NA150184 and from the National Research Foundation of South Africa (grant No. 103424).

\appendix

\section{Physical Models for the Observed Cross Coupling}
\label{sec:cross_coupling_models}
Here we explore the feasibility for some simple physical models as the origin of the ``high-delay'' cross coupling systematics investigated in \autoref{sec:cross_coupling}.
In summary, we cannot find a single model that explains all of the observed behavior of the systematics, but we can tentatively rule out some simplistic models.
In what follows, we adopt the mathematical conventions in Section 2 of \citet{Kern2019a} when discussing voltage spectra, visibilities and coupling coefficients.
Specifically, for two antennas 1 \& 2 with intrinsic voltage spectra $v_1$ and $v_2$, we can write the voltage of antenna 1 corrupted by a cable reflection as
\begin{align}
v^\prime_1 = v_1(1 + \epsilon_{11})
\end{align}
where $\epsilon_{11}$ is the cable reflection coefficient, and we can write the voltage of antenna 1 corrupted by cross coupling from antenna 2 as
\begin{align}
v^\prime_1 = v_1 + \epsilon_{21}v_2
\end{align}
where $\epsilon_{21}$ is the cross coupling of antenna 2's voltage into antenna 1's voltage.
If the uncorrupted cross-correlation visibility is $V_{12} = v_1v_2^\ast$, then the visibility corrupted by a reflection from antenna 1 can be written as
\begin{align}
V^\prime_{12} = v^\prime_1 v_2^\ast = v_1v_2^\ast (1 + \epsilon_{11}),
\end{align}
and the visibility corrupted by cross coupling can be written as
\begin{align}
V^\prime_{12} = v^\prime_1v_2^\ast = v_1v_2^\ast + \epsilon_{21}v_2v_2^\ast.
\end{align}

\subsection{A Noise Source in the Field}
A cross coupling-like signal can be generated by a stable noise source in the field, which will not fringe over time.
We can rule this out as the systematic mechanism simply based on the fact that we observe cross coupling systematics on short baselines at delays of $\ge1000$ ns: any (unreflected) source situated in the field will have a maximum achievable delay corresponding to the baseline's geometric horizon, which for short baselines is from 50 - 150 ns.

\subsection{Mutual Coupling Boosted by Cable Reflections}
One way to get cross coupling at high delays is to take cross coupling at low delays (e.g. mutual coupling) and boost it to high delays via a cable reflection.
If antenna 1 observes cross coupling from antenna 2 that then travels down and gets reflected in the cables of antenna 1, we can write the final measured visibility as
\begin{align}
\label{eqn:cc1}
V^\prime_{12} = (v_1 + \epsilon_{21}v_2)(1 + \epsilon_{11})v_2^\ast = v_1v_2^\ast + v_1v_2^\ast\epsilon_{11} + \epsilon_{21}v_2v_2^\ast + \epsilon_{21}v_2v_2^\ast\epsilon_{11}.
\end{align}
On the RHS of \autoref{eqn:cc1}, we recognize the first term as the uncorrupted visibility, the second term as the cable-reflected visibility, the third term as the first-order cross coupling systematic at low delay, and the last term as the cross coupling systematic boosted to high delay.
What we find is that the systematic can only be boosted to specific delays, determined by the product $\epsilon_{21}\epsilon_{11}$.
What we see in the data (specifically the right side of \autoref{fig:hera_cross_corr_frate_dly}) are cross coupling systematics at delays that are not consistent with this expectation.
Furthermore, the high-delay systematics do not look like a shifted version of the low-delay systematics, which is also a prediction of this model.

\subsection{A Broadcasting Antenna}
This model is a hybrid of the first two models, and states that a single antenna, say antenna 3, receives sky signal that traverses down its signal chain, is reflected back up one of its cables, is \emph{re-broadcasted out into the field} and then picked-up by neighboring antennas, mimicking a stable noise source in the field that has acquired a large delay lag due to the cable reflection in antenna 3's signal chain.
We can write the visibility between antenna 1 and 2 in the presence of this signal as
\begin{align}
V_{12}^\prime &= (v_1 + \epsilon_{31}\epsilon_{33}v_3)(v_2 + \epsilon_{32}\epsilon_{33}v_3)^\ast \\
&= v_1v_2^\ast + v_1\epsilon_{32}^\ast\epsilon_{33}^\ast v_3^\ast + \epsilon_{31}\epsilon_{33}v_3v_2^\ast + \epsilon_{31}\epsilon_{33}v_{3}\epsilon_{32}^\ast\epsilon_{33}^\ast v_3^\ast.
\end{align}
We recognize the first term on the RHS as the uncorrupted visibility, the fourth term as a standard cross-coupling term (due to the auto-correlation nature of $v_3v_3^\ast$) that has had its large delay canceled out due to $\epsilon_{33}\epsilon_{33}^\ast$ and thus does not appear at high delays.
Only the second and third terms will appear at high delays, but we can see that these terms are actually fringing terms because they contain products like $v_1v_3^\ast$ rather than $v_3v_3^\ast$ and thus will not appear centered at a fringe rate of 0 mHz, as we observe in the data.

\subsection{Summary}
While we have tentatively ruled out a few simple physical models, we still cannot point to a single mode that seems to explain the wide variety of behavior observed in the high-delay systematics.
What we can say is that the high-delay $f\sim0$ mHz terms seem to be physically disconnected from the $f\sim0$ mHz terms at low delays (i.e. at each baseline's geometric horizon).
Regardless of its origin, we do know that the high-delay features do not look like an EoR signal, and can therefore  be filtered out of the data.
Work is currently underway to assess whether these systematics appear in the upgraded HERA Phase II system, and if so what can be done in the field to mitigate their presence.

\section{Software}
\label{appendix:software}
The analysis presented in this work relies heavily on the Python programming language (\url{https://www.python.org}), and Python software developed by HERA collaboration members.
Here we provide a list of these packages and their version or Git hash:
\texttt{aipy [v2.1.12]} (\url{https://github.com/HERA-Team/aipy}), \texttt{healvis [v1.0.0]} \citep[\url{https://github.com/RadioAstronomySoftwareGroup/healvis};][]{Lanman2019}, \texttt{hera\_cal [v2.0]} (\url{https://github.com/HERA-Team/hera_cal}), \texttt{hera\_sim [v0.0.1]} (\url{https://github.com/HERA-Team/hera_sim}), \texttt{pyuvdata [v1.3.6]} \citep[\url{https://github.com/RadioAstronomySoftwareGroup/pyuvdata};][]{Hazelton2017}, and \texttt{uvtools [v0.1.0]} (\url{https://github.com/HERA-Team/uvtools}).
These packages in turn rely heavily on other publicly available software packages, including:
\texttt{astropy [v2.0.14]} \citep[\url{https://astropy.org};][]{Astropy2013}, \texttt{healpy [v1.12.9]} (\url{https://github.com/healpy/healpy}), \texttt{h5py [v2.8.0]} (\url{https://www.h5py.org/}), \texttt{matplotlib [v2.2.4]} (\url{https://matplotlib.org}), \texttt{numpy [v1.16.2]} (\url{https://www.numpy.org}), \texttt{scipy [v1.2.1]} (\url{https://scipy.org}), and \texttt{scikit-learn [v0.19.2]} (\url{https://scikit-learn.org}).

\bibliographystyle{apj}
\bibliography{systematic_modeling}

\end{document}